\documentclass[twocolumn,prb,superscriptaddress,showpacs]{revtex4-1}

\usepackage{amsmath}
\usepackage{graphicx}
\usepackage{lipsum}
\usepackage{hyperref}
\begin{document}

\title{Anisotropic damping and wavevector dependent susceptibility of the spin fluctuations in La$_{2-x}$Sr$_{x}$CuO$_4$ studied by resonant inelastic x-ray scattering}
\author{H. C. Robarts} 
\affiliation{H. H. Wills Physics Laboratory, University of Bristol, Bristol BS8 1TL, United Kingdom}
\affiliation{Diamond Light Source, Harwell Science \& Innovation Campus, Didcot, Oxfordshire OX11 0DE, United Kingdom}
\author{M. Barth\'elemy}
\affiliation{H. H. Wills Physics Laboratory, University of Bristol, Bristol BS8 1TL, United Kingdom}
\author{K. Kummer}
\affiliation{European Synchrotron Radiation Facility, 71 Avenue des Martyrs, Grenoble, France}
\author{M. Garc\'ia-Fern\'andez}
\affiliation{Diamond Light Source, Harwell Science \& Innovation Campus, Didcot, Oxfordshire OX11 0DE, United Kingdom}
\author{J.\;Li}
\affiliation{Diamond Light Source, Harwell Science \& Innovation Campus, Didcot, Oxfordshire OX11 0DE, United Kingdom}
\affiliation{Beijing National Laboratory for Condensed Matter Physics and Institute of Physics, Chinese Academy of Sciences, Beijing 100190, China}
\author{A. Nag}
\author{A. C. Walters}
\author{K. J. Zhou}
\email{kejin.zhou@diamond.ac.uk}
\affiliation{Diamond Light Source, Harwell Science \& Innovation Campus, Didcot, Oxfordshire OX11 0DE, United Kingdom}
\author{S. M. Hayden}
\email{s.hayden@bristol.ac.uk}
\affiliation{H. H. Wills Physics Laboratory, University of Bristol, Bristol BS8 1TL, United Kingdom}

\begin{abstract}
We report high-resolution resonant inelastic x-ray scattering (RIXS) measurements of the collective spin fluctuations in three compositions of the superconducting cuprate system La$_{2-x}$Sr$_{x}$CuO$_4$. We have mapped out the excitations throughout much of the 2-D $(h,k)$ Brillouin zone. The spin fluctuations in La$_{2-x}$Sr$_{x}$CuO$_4$ are found to be fairly well-described by a damped harmonic oscillator model, thus our data allows us to determine the full wavevector dependence of the damping parameter. This parameter increases with doping and is largest along the ($h, h$) line, where it is peaked near $(0.2, 0.2)$. We have used a new procedure to determine the absolute wavevector-dependent susceptibility for the doped compositions La$_{2-x}$Sr$_{x}$CuO$_4\;(x=0.12, 0.16)$ by normalising our data to La$_2$CuO$_4$ measurements made with inelastic neutron scattering (INS). We find that the evolution with doping of the intensity of high-energy excitations measured by RIXS and INS is consistent. For the doped compositions, the wavevector-dependent susceptibility is much larger at $(\frac{1}{4},\frac{1}{4})$ than at $(\frac{1}{2},0)$. It increases rapidly along the $(h,h)$ line towards the antiferromagnetic wavevector of the parent compound $(\frac{1}{2},\frac{1}{2})$. Thus, the strongest magnetic excitations, and those predicted to favour superconductive pairing, occur towards the $(\frac{1}{2},\frac{1}{2})$ position as observed by INS. 
\end{abstract}

\maketitle

\section{Introduction}
The origin of high temperature superconductivity (HTS) in doped layered cuprate materials remains a subject of intense interest in both experimental and theoretical research, despite over 30 years of activity. It is widely believed that the magnetic degrees of freedom and in particular spin fluctuations are primarily responsible for superconductive pairing in the cuprates \cite{chubukov2003, scalapino2012, eschrig2006, keimer2015}. In this case, it is important to characterize the collective spin excitations as a function of wavevector, energy, doping and temperature to see how they correlate with the occurrence of superconductivity and compare with theoretical models.  
%Experimental studies on cuprate materials have variously revealed information about the magnetic interactions and electronic structure. By focusing on the interactions in the CuO$_4$ planes, two main classes of theory have emerged to deal with these findings \cite{chubukov2002}. The first, considers holes introduced to the Mott insulating phase hopping to neighbouring sites with amplitude $t$ to minimise kinetic energy. This is balanced by the superexchange interaction $J$ between Cu atoms \cite{tohyama2011,ogata2008,lee2006}. The second, considers electrons from near the Fermi surface coupling to magnetic excitations \cite{chubukov2002}. In this model, magnetic excitations replace the phonons which facilitate pairing in conventional superconductivity. Both theories rely on the proximity of superconductivity to a magnetic phase as is the case in cuprates. Resolving details across the cuprate phase diagram therefore remains essential for understanding these mechanisms \cite{scalapino2012, eschrig2006, keimer2015} and particular interest is focused on the relationship between magnetic excitations and the emergence of superconductivity. Detailed mapping of both the evolution and nature of collective excitations is therefore desirable. 

Resonant inelastic x-ray scattering (RIXS) \cite{ament2011,braicovich2009,braicovich2010,LeTacon2011_LGCS,Ghiringhelli2012_GLMB,dean2012,dean2013,dean2015,monney2016,chaix2018} and inelastic neutron scattering (INS) \cite{Hayden1991_HAOT, hayden1996, Arai1999_ANEE, Dai1999_DMHA, coldea2001, headings2010, lipscombe2007} are complementary probes which directly yield information about the wavevector and energy of the dynamical structure factor $S(\mathbf{Q},\omega)$ or dynamic susceptibility (response function) $\chi^{\prime\prime}(\mathbf{Q},\omega)$ at high frequencies. The La$_{2-x}$Sr$_x$CuO$_4$ (LSCO) system allows the evolution of $S(\mathbf{Q},\omega)$ to be measured across the phase diagram, from the antiferromagnetic (AF) parent compound La$_2$CuO$_4$ (LCO) through superconducting compositions. 

In La$_2$CuO$_4$, the spin waves have their lowest energies at the $\Gamma$, $\mathbf{Q}$=($0, 0$) and $M$, $\mathbf{Q}$=($\frac{1}{2}, \frac{1}{2}$) positions and $\chi^{\prime\prime}(\mathbf{Q},\omega)$ is small near $\Gamma$ and largest near $M$. INS measurements\cite{Hayden1991_HAOT, coldea2001, headings2010} throughout the Brillouin zone have shown that the magnetic excitations can be fairly well-described as spin waves derived from a Heisenberg model with next-nearest neighbour interactions including a ring exchange.  As expected, they are strongest near the AF wavevector $\mathbf{Q}$=($\frac{1}{2}, \frac{1}{2}$) and show anomalously strong damping at the $X$ or ($\frac{1}{2},0$) position \cite{headings2010,dean2012,Sandvik2001_SaSi}.  

For superconducting compositions in LSCO, INS shows that the strongest response \cite{hayden1996,vignolle2007,Wakimoto2007_WYTF, lipscombe2007, lipscombe2009} occurs near $\mathbf{Q}$=($\frac{1}{2}, \frac{1}{2}$) at low and intermediate energies (0--150\;meV), with comparable intensity to the parent antiferromagnet.  For optimally doped ($x=0.16$) LSCO, an incommensurate structure is observed\cite{vignolle2007} for $\hbar \omega \lesssim 25$\;meV. Above 50\;meV the magnetic excitations disperse\cite{vignolle2007, lipscombe2007, lipscombe2009} away from ($\frac{1}{2}, \frac{1}{2}$).  At high energies, $\hbar \omega \approx 250$\;meV, excitations are observed\cite{hayden1996} on the Brillouin zone boundary at $\mathbf{Q}=(\frac{1}{2},0)$ in LSCO ($x=0.14$) demonstrating the persistence of high energy spin excitations for superconducting compositions. For overdoped compositions\cite{Wakimoto2007_WYTF, lipscombe2007} $x=0.22-0.25$, the lower energy ($\hbar \omega \sim 50$\;meV) features observed at optimal doping are suppressed.

\begin{figure}
\centering
\includegraphics[width=\linewidth]{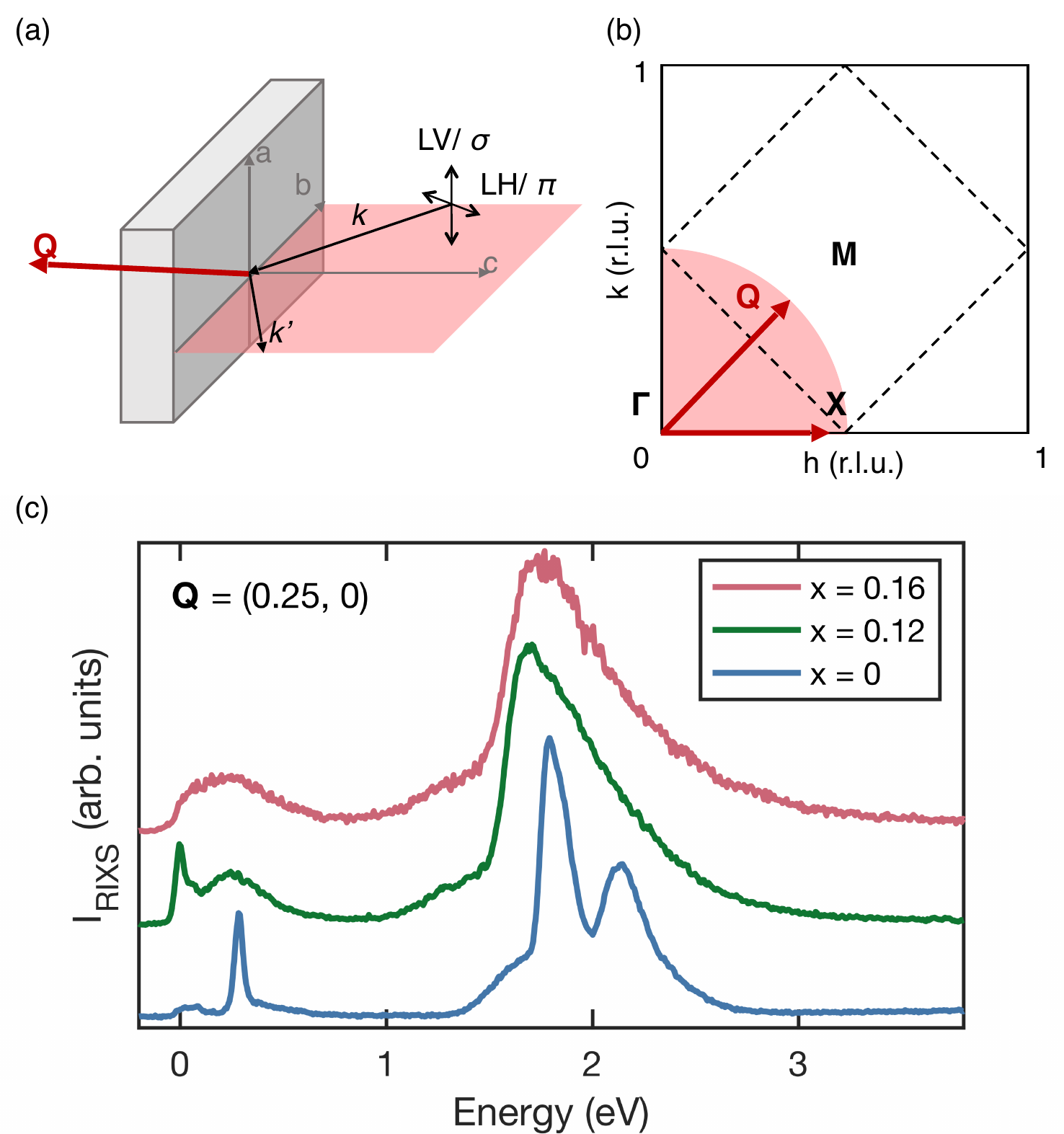}
\caption{Experimental geometry showing (a) the scattering plane in relation to the crystal axis and (b) the resulting measured region of the LSCO unit cell. The measured wave-vectors \textbf{Q} = $\textbf{k}$ - $\textbf{k}^{\prime}$ giving the measurement regions indicated in red. With our definitions, grazing-out for $\textbf{k}^{\prime}$ corresponds to positive $h$ and $k$. (c) shows example RIXS spectra from each compound at \textbf{Q} = ($0.25, 0$).}
\label{fig:intro}
\end{figure}

Cu $L_3$ RIXS\cite{braicovich2009,braicovich2010,dean2012,dean2013,dean2015,monney2016,meyers2017,chaix2018} measurements of the spin fluctuation in LSCO are complementary to INS. They are restricted to a circular region in $(h,k)$ centered on $\Gamma$ [see Fig.~\ref{fig:intro} (a) and (b) ] but are able to isolate high energy excitations ($\hbar\omega\gtrsim$ 300\;meV) more easily. Early RIXS measurements in LSCO \cite{braicovich2010} verified the existence of dispersing spin fluctuations. Spin excitations are observed\cite{braicovich2010,dean2013,dean2015,monney2016,meyers2017,chaix2018} throughout the first AF Brillouin zone including at the boundary [e.g. ($\frac{1}{2},0$) position] where INS\cite{hayden1996} also finds excitations. 
RIXS studies suggest that these excitations show wavevector-dependent damping \cite{dean2013,monney2016,meyers2017}. Spin fluctuations persist to overdoped compositions and evolve relatively slowly with doping \cite{dean2013,monney2016}.

The improved energy resolution of the measurements we have performed allows us to model the nature of the spin fluctuations more precisely. The motivation of this work is to perform a systematic characterisation of the spin fluctuations in LSCO with this enhanced energy resolution including mapping the \textbf{Q}-dependence of the frequency and damping throughout a 2-D portion of the Brillouin zone. We also aim to bridge the techniques of INS and RIXS to establish an estimate of the absolute spin susceptibility.

Here we report RIXS measurements on three dopings of LSCO, $x$ = 0, 0.12 and 0.16. We have made use of the high resolution and high intensity of the RIXS spectrometers ID32 at the European Synchrotron Radiation Facility (ESRF) and I21 at the Diamond Light Source (DLS) to map out magnetic spectra over 2-D $(h,k)$ space. We find that, for doped compositions, the magnetic response is fairly well-described by a damped harmonic oscillator line shape. The pole frequency and damping are strongly anisotropic in agreement with previous studies along the $(h,0)$\cite{dean2013} and $(h,h)$\cite{meyers2017} lines, with the strongest damping along the $(h,h)$ line and centred near (0.2, 0.2) for the optimally doped composition. By comparing data on La$_2$CuO$_4$, where the spin waves are well studied, with LSCO, we make quantitative estimates of the wavevector-dependent susceptibility $\chi^{\prime}(\mathbf{Q})$. This quantity is a vital input to theories of the HTS phenomenon \cite{chubukov2003, scalapino2012, eschrig2006}.

\section{Experimental details}
\label{sec:methods_experiment}
\subsection{Samples}
Measurements were performed on single crystal samples of LSCO. Three different compositions were measured: the $x$ = 0 parent compound which displays antiferromagnetism below $T_N \simeq$ 320\;K \cite{headings2010} and two hole-doped compounds\cite{croft2014,vignolle2007}, $x$ = $0.12 \pm 0.005$ ($T_c=29.5$\;K) and $x=0.16 \pm 0.005$ ($T_c=38$\;K). The $x$ = 0.16 composition is close to optimal doping for the superconducting phase and $x$ = 0.12 shows charge density wave (CDW) order with short range CDW correlations developing at $T \sim 150$\;K and a longer range CDW developing at $T_{\text{CDW}} \simeq$ 75\;K \cite{croft2014}. Crystals were grown via the travelling solvent floating zone technique and used in previous neutron \cite{headings2010,vignolle2007,lipscombe2007}  and x-ray \cite{croft2014} studies. The crystals were re-cut into posts with typical dimensions $\simeq$ 2 x 1 x 1 mm$^3$. The samples were aligned using Laue x-ray diffraction and cleaved \textit{in-situ} to expose a clean surface to the beam. The sample used for measurements of the $x$ = 0.12 compound at the ESRF was polished following the procedure in Croft \textit{et al.} \cite{croft2014}. For the same composition, the elastic peak observed close to the specular condition is approximately 15 times greater in the polished sample compared to the cleaved sample. This makes the low energy excitations at low \textbf{Q} difficult to extract and we therefore only use data from this sample in the map plots. We verified that the lineshape, intensity and energy of the magnetic excitations is the same in both datasets.

\subsection{Notation}
LSCO undergoes a structural transition to a low-temperature orthorhombic (LTO) phase below $T_{\text{LTO}} \simeq$ 240\;K, however, we use the high-temperature tetragonal (HTT) I/4mmm crystal structure notation to allow comparison between the three compounds. In this notation, $a$ = $b \simeq$ 3.8\;\AA, $c \simeq$ 13.2\;\AA. The momentum transfer $\mathbf{Q}$ is defined in reciprocal lattice units (r.l.u.) as $\mathbf{Q} = h\mathbf{a^{*}} + k\mathbf{b^{*}} + l\mathbf{c^{*}}$ where $\mathbf{a^{*}} = 2\pi/a$ etc. The measured excitations are labelled via their energies $\hbar \omega =c\left|\mathbf{k}\right| - c\left|\mathbf{k}^{\prime}\right|$ and momenta $\mathbf{Q} = \mathbf{k} - \mathbf{k}^{\prime}$, where  $\mathbf{k}$ and $\mathbf{k}^{\prime}$ are in initial and final wavevectors.

\subsection{Spectrometers}
High resolution RIXS spectra were measured at beamline ID32 of the ESRF \cite{braicovich2012,brookes2018} and the I21 RIXS spectrometer at DLS\cite{diamondi21}. The incoming beam energy was tuned to the Cu L$_3$-edge ($\sim$ 932\;eV) with linear horizontal (LH) $\pi$ polarisation. We present LH data from the grazing-out orientation where the single magnon intensity is favoured \cite{braicovich2010b,moretti2011}. Recent experiments with polarisation analysis\cite{peng2018, fumagalli2019} have established that this configuration is primarily sensitive to magnetic scattering.    Samples were mounted on the sample holder in ultra-high vacuum and cooled to $T \simeq 20$\;K. Magnetic excitations in cuprates are dispersive predominantly in the $a$-$b$ plane of LSCO, allowing paths to be measured in the ($h,k$) plane by varying the sample orientation, and keeping the scattering angle $2 \theta$ fixed at 146$^{\circ}$ and 149.5$^{\circ}$ for I21 and ID32 respectively. The scattering geometry is shown in Fig. \ref{fig:intro} (a). We assume there is negligible dispersion in the features of interest from variation of $l$, and therefore we focus only on the momentum transferred in the ($h,k$) plane. Spectra were principally measured along the two high-symmetry lines ($h, 0$) and ($h,h$) as indicated with red arrows in Fig. \ref{fig:intro} (b) with energy resolution $\Delta E$ $\simeq$ 35\;meV.  The $x$ = 0 and 0.12 measurements were performed at I21 and the $x$ = 0.16 measurements were performed at ID32 and repeated at I21. In both doped compounds, further measurements were performed at ID32 with $\Delta$E $\simeq$ 50\;meV on a grid of \textbf{Q}-points evenly distributed throughout a quadrant of the Brillouin zone indicated by the red shaded region in Fig. \ref{fig:intro} (b). The energy resolution was established using elastic scattering from a silver paint or carbon tape reference. For I21, a background was measured from either a dark image taken after the collection or by fitting a constant background outside the excitation range, $\leq -0.1$\;eV and $\geq$ 5\;eV. 
\subsection{Analysis}
\label{sec:methods_analysis}
\subsubsection{Data processing}
In order to carry out a quantitative analysis of the data, we follow recent practice \cite{braicovich2009, braicovich2010, dean2013b, monney2016, meyers2017, peng2018} and assume that the magnetic intensity observed in RIXS is proportional to the spin-spin dynamical structure factor $S(\mathbf{Q},\omega)$ which is used to interpret neutron scattering experiments \cite{Lovesey1986_Love}. $S(\mathbf{Q},\omega)$ is, in turn, proportional to $\chi^{\prime\prime}(\mathbf{Q},\omega)$ multiplied by the Bose factor $n(\omega)+1 = [1-\exp(-\hbar\omega/k_B T)]^{-1}$. Clearly, the scattering processes in RIXS and INS are very different, with the observed RIXS intensity being dependent on the relative orientation of the photon electric field to the Cu $3d$ orbitals as well as the absorption of the x-ray photons within the sample.  These factors are known to vary slowly with $\mathbf{Q}$ \cite{ament2009, Jia2014_JNWK}, nevertheless, to correct for these effects we initially normalise our raw counts $I_{\mathrm{raw}}$ to the energy-integrated $dd$ excitation intensity obtained from the same spectrum. The intensity of the $dd$ excitations is known to be dependent on the polarisation $\epsilon$ and wavevector $\mathbf{k}$ and can be described by a function $g(\epsilon,\epsilon^{\prime},\mathbf{k},\mathbf{k^{\prime}})$. We denote the measured intensity $I_{\mathrm{RIXS}}$ as $I_{\mathrm{raw}}/g $ where $g = \int g(\epsilon,\epsilon^{\prime},\mathbf{k},\mathbf{k^{\prime}})\,d\omega$ is the integral described above evaluated over the range 1--3\;eV.

The spectra were aligned to the elastic reference and the exact zero-energy position was established by fitting an elastic peak with a Gaussian function. The aligned spectra were modelled within a range --80 to 800\;meV. As well as the spin excitations,  we fit an elastic peak and low-energy excitations, which are interpreted as phonons, using Gaussian functions. Electron-hole excitations and broadened $dd$ excitations contribute to the low-energy RIXS scattering for doped compositions \cite{dean2013}. This contribution was modelled with a linear function which was fixed for all spectra of the same composition. The gradient of the linear function was found by fitting the spectra at low \textbf{Q}. In the insulating parent compound this contribution was not required. However, a broad continuum of multimagnon excitations is resolvable at $\sim$ 400--600\;meV. This was modelled with a Gaussian function.

The spectra were not deconvolved to take account of the instrument energy resolution $\simeq 35$\;meV. The most noticeable effect of this was in the determination of $\gamma$ and $\Gamma$ values (see Sec.\;\ref{sec:DHO}). We estimate that our fitted values are increased by 5\% in the worse case.

\begin{figure}
\begin{center}
\includegraphics[width=\linewidth]{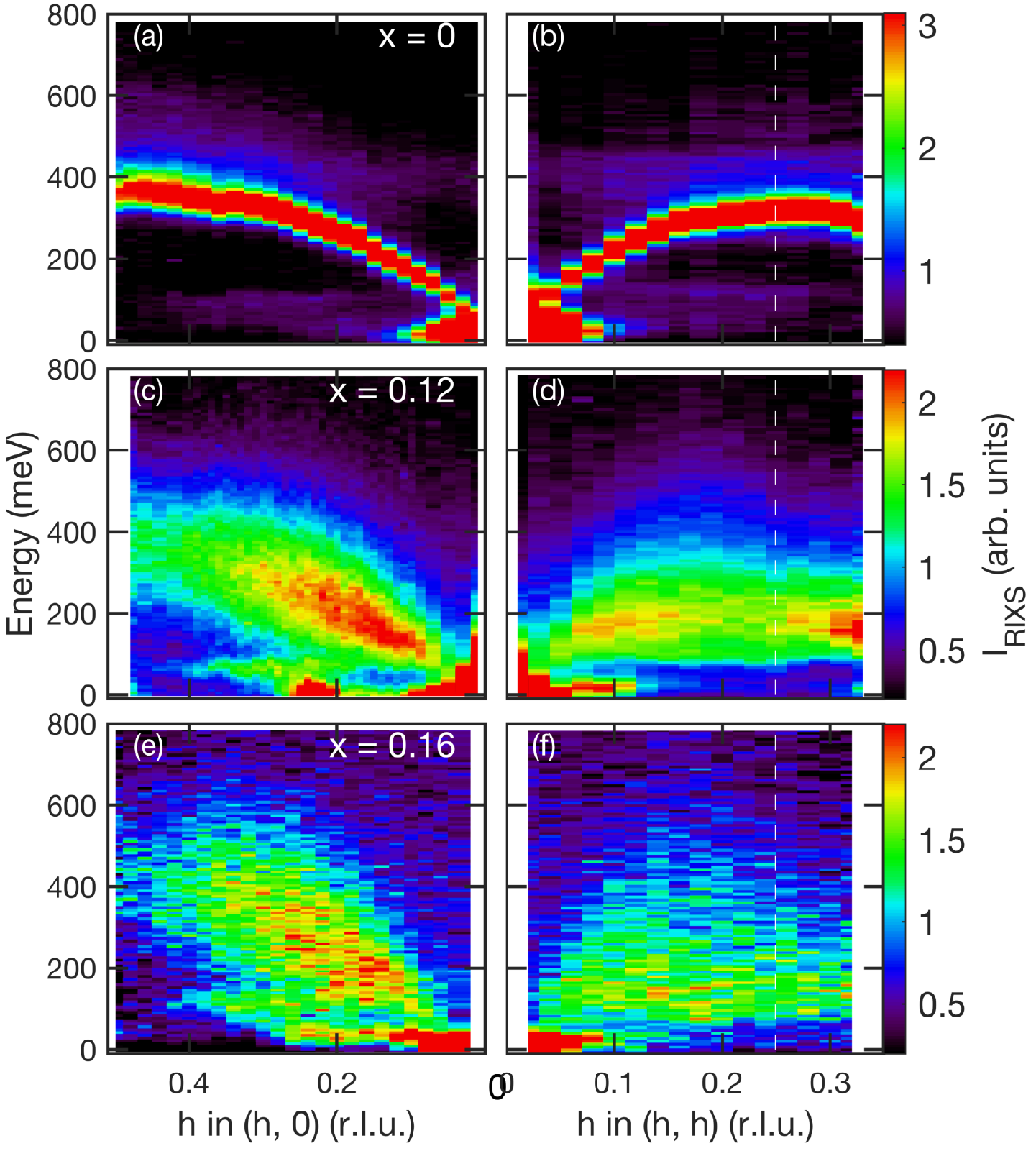}
\end{center}
\caption{$I_{\textrm{RIXS}}$ intensity maps as a function of $\textbf{Q}$ in LSCO $x$ = 0 ($T \approx 20$\;K), 0.12 and 0.16 ($T \approx 30$\;K). Showing measurements along the ($h, 0$) and ($h, h$) lines. The measurements were performed in grazing-out geometry and with LH polarization at I21 at Diamond Light Source. The configuration favours magnetic scattering. All three compositions show charge scattering in the form of phonons below 100\;meV and a charge density wave peak is observed near $h=0.23$ in $x=0.12$. The dashed white line marks the antiferromagnetic Brillouin zone boundary (see Fig.\;\ref{fig:intro}).} 
\label{fig:intensity_maps}
\end{figure} 

\begin{figure*}
\centering
\includegraphics[width=\linewidth]{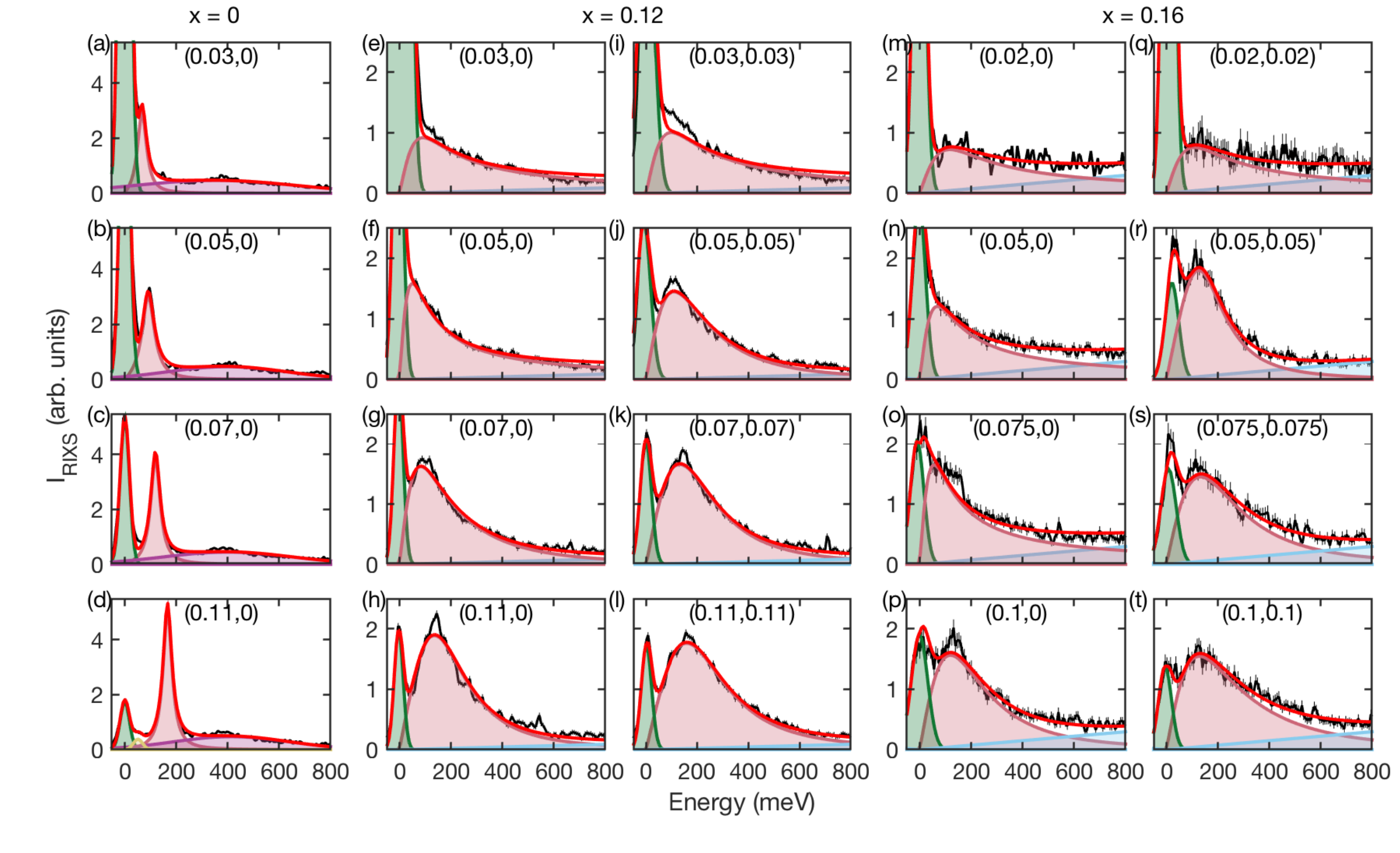}
\caption{Examples of fitted RIXS spectra from LCO and LSCO $x$ = 0.12 (performed at I21 at DLS) and $x$ = 0.16 (performed at ID32 at the ESRF). Showing data in the low \textbf{Q} regime from high symmetry directions ($h, 0$), ($h, h$). The data have intensity $I_{\textrm{RIXS}}$ indicating that they are normalised to an integration over the range of the $dd$ excitations, $g$. The total fit to the data is indicated in red, the magnetic excitations in pink, elastic peak in green, multi-magnons in purple and background in blue. Panels (e), (f), (i), (j), (m), (n), (o) and (q) are fitted with the ODHO function.}
\label{fig:fit_plots_lowQ}
\end{figure*} 

\subsubsection{Damped harmonic oscillator model}
\label{sec:DHO}
A damped harmonic oscillator (DHO) model may be used to describe a given spin-wave mode with wave vector \textbf{Q}. This approach has recently been taken in a number of RIXS studies \cite{dean2013,monney2016,lamsal2016,peng2018}. The analogous mechanical DHO equation is \cite{chaikin1995}
\begin{equation}
\ddot{x} + \omega_0^2x + \gamma \dot{x} = f/m,
\label{equ:DHO}
\end{equation}
where $\omega_0$ is the frequency of the undamped mode and $\gamma$ is the damping parameter. In our case, both of these are \textbf{Q}-dependent, thus $\omega_0 = \omega_0(\mathbf{Q)}$ and $\gamma = \gamma(\mathbf{Q)}$.

\begin{figure*}
\centering
\includegraphics[width=\linewidth]{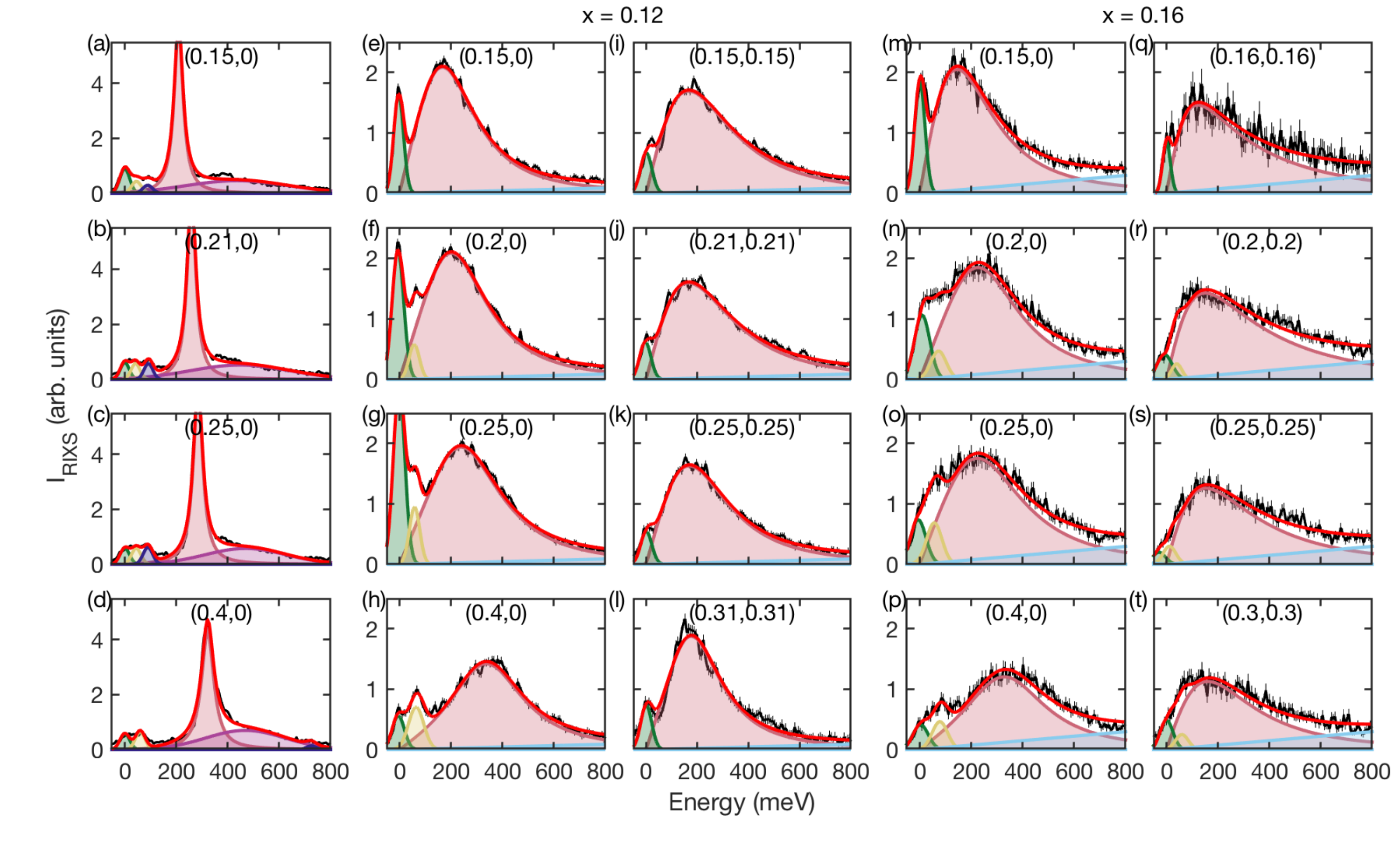}
\caption{Examples of fitted RIXS spectra from LCO and LSCO $x$ = 0.12 (performed at I21 at DLS) and $x$ = 0.16 (performed at ID32 at the ESRF). Showing data in the high \textbf{Q} regime from high symmetry directions ($h, 0$), ($h, h$). The data have intensity $I_{\textrm{RIXS}}$ indicating that they are normalised to an integration over the range of the $dd$ excitations, $g$. The total fit to the data is indicated in red, the DHO magnetic excitations in pink, elastic peak in green, phonon excitations in yellow and dark blue, multi-magnons in purple and background in light blue.}
\label{fig:fit_plots}
\end{figure*}

The imaginary part of the DHO response function for a given wavevector can be written as,
\begin{equation}
\label{equ:damping}
\chi^{\prime\prime}(\mathbf{Q}, \omega) = \frac{\chi^{\prime}(\textbf{Q}) \, \omega_0^2(\mathbf{Q)} \, \gamma(\mathbf{Q)}\, \omega}{\left[\omega^2 -\omega_0^2(\mathbf{Q)}\right]^2+\omega^2\gamma^2(\mathbf{Q)}},
\end{equation}
where $\chi(\mathbf{Q}) \equiv \chi^{\prime}(\mathbf{Q}) \equiv \chi^{\prime}(\mathbf{Q},\omega=0)$ is the real part of the zero frequency susceptibility. The solution of Eq.~\ref{equ:DHO} can be represented by two poles with complex frequencies:
\begin{equation}
\omega = \pm \left[\omega_0^2 - \left(\gamma^2/4\right)\right]^{\frac{1}{2}} = \pm \omega_1 - \frac{i\gamma}{2}.
\end{equation}
If $\omega_0^2 \geq \gamma^2/4$, $\omega_1$ is real and the frequency of the pole. The solutions (response) correspond to damped oscillations in time. If $\omega_0^2 \leq \gamma^2/4$, $\omega_1$ is imaginary and the system is overdamped. We may introduce a third frequency, $\omega_{\textrm{max}}$, defined as the frequency at the peak in $\chi^{\prime\prime}(\omega)$. This can be shown to be
\begin{equation}
\omega_{\textrm{max}}=\frac{1}{6}\sqrt{12\omega_0^2-6\gamma^2+6\sqrt{\gamma^4-4\gamma^2\omega_0^2+16\omega_0^4}}.
\end{equation}

Using the DHO function (Eqn.\;\ref{equ:damping}) to analyse all of the data allows a consistent model to be applied to the underdamped and overdamped regimes. This is useful when comparing excitations from undoped and doped compositions. In particular, $\gamma/2 > \omega_0$ is allowed in this model, however, beyond critical damping, $\gamma/2 = \omega_0$, the shape of the response function evolves relatively slowly. Further, the fitted values of $\gamma$ and $\omega_0$ become correlated. This is the case at small $|\mathbf{Q}|$. In the limit of large damping\cite{chaikin1995} $\gamma/\omega_0 \rightarrow \infty$, $\chi^{\prime\prime}(\mathbf{Q},\omega)$ can be approximated by the overdamped harmonic oscillator (ODHO) Lorentzian form,
\begin{equation}
\chi^{\prime\prime} (\mathbf{Q}, \omega) = \frac{\chi^{\prime}(\mathbf{Q})\Gamma(\mathbf{Q}) \omega}{\omega^2 + \Gamma^2(\mathbf{Q})}.
\label{eqn:damping_low_Q}
\end{equation}
Eqn.\;\ref{eqn:damping_low_Q} only has two parameters, $\chi^{\prime}$ and the relaxation rate $\Gamma = \omega_0^2/\gamma$.  We found it convenient to use Eqn.\;\ref{eqn:damping_low_Q} in some of the overdamped region. Thus, the grey region in Figs.\;\ref{fig:fit_parameters} and \ref{fig:chi_fit} indicate the low \textbf{Q} regime where Eqn.\;\ref{eqn:damping_low_Q} is used to fit the data.

\section{Results}
\subsection{RIXS spectra of La$_{2-x}$Sr$_x$CuO$_4$}

Fig. \ref{fig:intro} (c) shows example spectra from each composition at \textbf{Q} = (0.25,0).  The low-energy magnetic spectrum of the parent ($x=0$) compound (bottom), is dominated by resolution-limited spin-wave excitations. The magnetic excitations in the doped $x=0.12$ (middle) and $x=0.16$ (top) compositions are considerably broader as noted in previous studies\cite{braicovich2010,Ghiringhelli2012_GLMB,dean2012,dean2013,dean2015,monney2016}. The \textit{dd} excitations occur in the energy range 1--3\;eV. These are considerably broadened and shifted to lower energy in the doped compositions. The spectra are consistent with published lower resolution data \cite{dean2013}.

\begin{figure*}
\centering
\includegraphics[width=\linewidth]{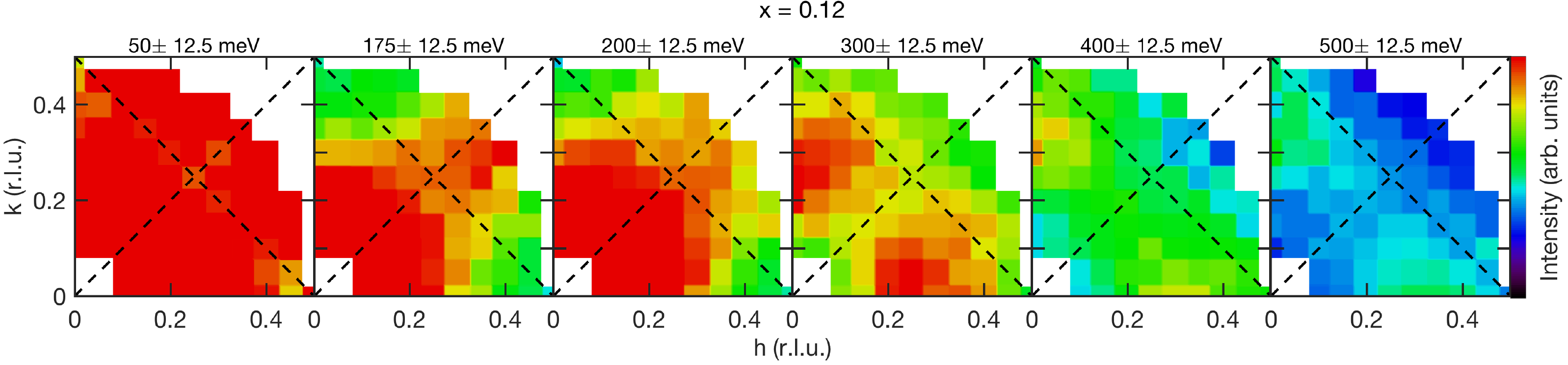}
\caption{Constant energy maps of RIXS intensity as a function of \textbf{Q}-vector ($h, k$) in LSCO $x$ = 0.12. The data have intensity $I_{\textrm{RIXS}}$ indicating that they are normalised to an integration over the range of the $dd$ excitations, $g$. Measurements were performed at ID32 at the ESRF with $\Delta$E $\simeq$ 50\;meV. Black dashed lines indicate the zone boundary and high symmetry directions.}
\label{fig:LSCO_maps}
\end{figure*}

Figs.~\ref{fig:fit_plots_lowQ} and \ref{fig:fit_plots} show examples of our RIXS data. Spectra such as those in Figs.\;\ref{fig:fit_plots_lowQ} and \ref{fig:fit_plots} are collected together into intensity maps plotted as a function of $\mathbf{Q}$ and energy in Fig.~\ref{fig:intensity_maps}.  Thus Fig.~\ref{fig:intensity_maps} gives an overall picture of the excitations observed in the present study. The strongest feature in Fig.~\ref{fig:intensity_maps}(a) is the magnon which disperses to an energy $\sim 355 \pm 34$\;meV along $(h,0)$ in agreement with previous studies\cite{headings2010,dean2012}. The magnetic excitations are much broader in energy for doped compositions as shown in Figs.~\ref{fig:intensity_maps}(c)-(f). Phonons can be seen in the La$_2$CuO$_4$ spectra below 100~meV, for example in Fig.~\ref{fig:fit_plots}(c) and also visible in the map plots in Fig.~\ref{fig:intensity_maps}. In Fig.~\ref{fig:intensity_maps}(c), for $x=0.12$ and $0.16$ a particularly strong phonon branch can be seen below 100\;meV along $(h,0)$ near $h=0.3$. This indicates coupling to charge excitations. In the $x=0.12$ composition, CDW order is seen near $h=0.23$. Similar behaviour\cite{chaix2017} is seen in Bi$_2$Sr$_2$CaCu$_2$O$_{8+\delta}$. 

In addition to the high-symmetry direction measurements shown in Figs.~\ref{fig:intensity_maps}-\ref{fig:fit_plots_lowQ}, a full quadrant of the Brillouin zone was examined by mapping ($h,k$) in the $x$ = 0.16 and 0.12 compounds. Approximately 90 spectra were collected at the ID32 beamline, distributed throughout the zone with spacing 0.05 (r.l.u.). The RIXS intensity is plotted as a function of ($h, k$) for several energy slices and for $x=0.12$ in Fig. \ref{fig:LSCO_maps} where areas of high-intensity correspond to the spin-excitation intensity. These measurements were performed with lower resolution ($\Delta E\simeq$ 50\;meV). The plots are smoothed by averaging neighbouring points within $|\Delta\mathbf{Q}|$ = 0.05\;r.l.u.. At low energies, the maximum in the RIXS intensity appears at low \textbf{Q} and is approximately symmetrically distributed around $\Gamma$. As the energy increases, peaks develop along $(h,0)$ and $(0,k)$ and move to larger $h$ and $k$. It is interesting to note that quite similar behaviour is observed\cite{headings2010} in La$_2$CuO$_4$, where $(h,k)$ maps measured with INS show a peak in the intensity at ($\frac{1}{2},0$) for energies above about 320\;meV. The maps show that for doped LSCO the magnetic spectral weight persists to higher energies near ($\frac{1}{2},0$) than in other parts of the Brillouin zone. This observation is consistent with previous work\cite{braicovich2010,dean2013,dean2015,monney2016,meyers2017}.  

\subsection{DHO fitting}
\label{sec:dho_fitting}
\begin{figure}[ht]
\centering
\includegraphics[width=\linewidth]{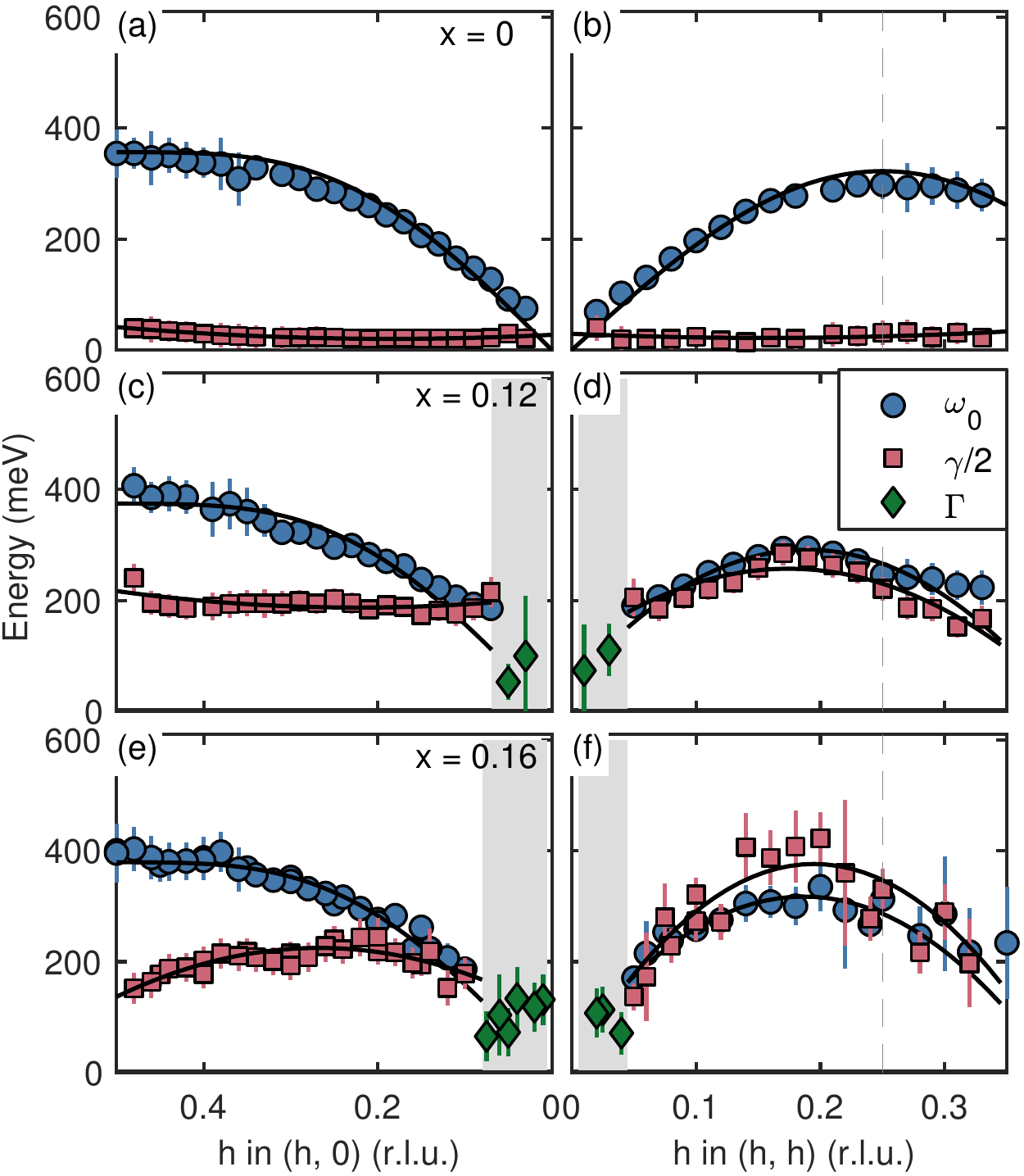}
\caption{Summary of fit parameters to a DHO response model as a function of wave-vector \textbf{Q} along high symmetry directions ($h, 0$) and ($h, h$). $\omega_{0}$ is indicated with blue circles and the damping coefficient $\gamma$/2 is shown as red squares. In the low \textbf{Q} regime the ODHO relaxation rate is given by $\Gamma$ which is shown as green diamonds. Errors are from fitting considering the standard error in the raw data. Solid lines are a cubic polynomial fit to the data. Data in panel (a), (b) and (c) all contain data measured at I21 and panel (c) contains additional data measured at ID32. The dashed grey line marks the AF Brillouin zone boundary. }
\label{fig:fit_parameters}
\end{figure}

Figs.~\ref{fig:fit_plots_lowQ} and \ref{fig:fit_plots} show fits of the damped harmonic oscillator (DHO) model (Sec.~\ref{sec:methods_analysis}) together with phonon peaks and background to the data.  The 35\;meV resolution of the instrument allows the phonons and elastic peaks to be separated from the DHO response. For example, in La$_2$CuO$_4$ the frequencies are approximately wavevector independent with energies $\sim 45$ and 90\;meV which are attributed to CuO bond-bending and bond-stretching modes respectively \cite{pintschovius2005,Reznik2012,devereaux2016}. As can be seen from the figures, the DHO model generally describes the magnetic excitations well. The measured spectra are shown in black with the total fitted function indicated in red with constituent functions below. The parameters $\omega_0$ and $\gamma/2$ extracted from DHO fits are plotted for \textbf{Q} = ($h, 0$) and ($h,h$) in Fig.~\ref{fig:fit_parameters} for each compound. Eqn.\;\ref{eqn:damping_low_Q} is used to fit the small \textbf{Q} (grey) regime and the resulting relaxation rate $\Gamma$ is shown. Hole doping the parent compound increases $\gamma/2$. In the doped compounds, it can be comparable to $\omega_0$. The damping is anisotropic in wavevector\cite{dean2013,dean2015,monney2016,meyers2017}, that is $\gamma/2$ is larger along ($h,h$) than along ($h, 0$). Our data also reveals that the anisotropy of the damping does not reflect the antiferromagnetic Brillouin zone as $\gamma/2$ peaks at approximately ($0.2, 0.2$) [rather than ($\frac{1}{4}, \frac{1}{4}$)] along ($h, h$). This effect can be seen both for $x=0.12$ and $x=0.16$ Fig.~\ref{fig:fit_parameters}(d,f). 

\begin{figure*}
\begin{center}
\includegraphics[width=\linewidth]{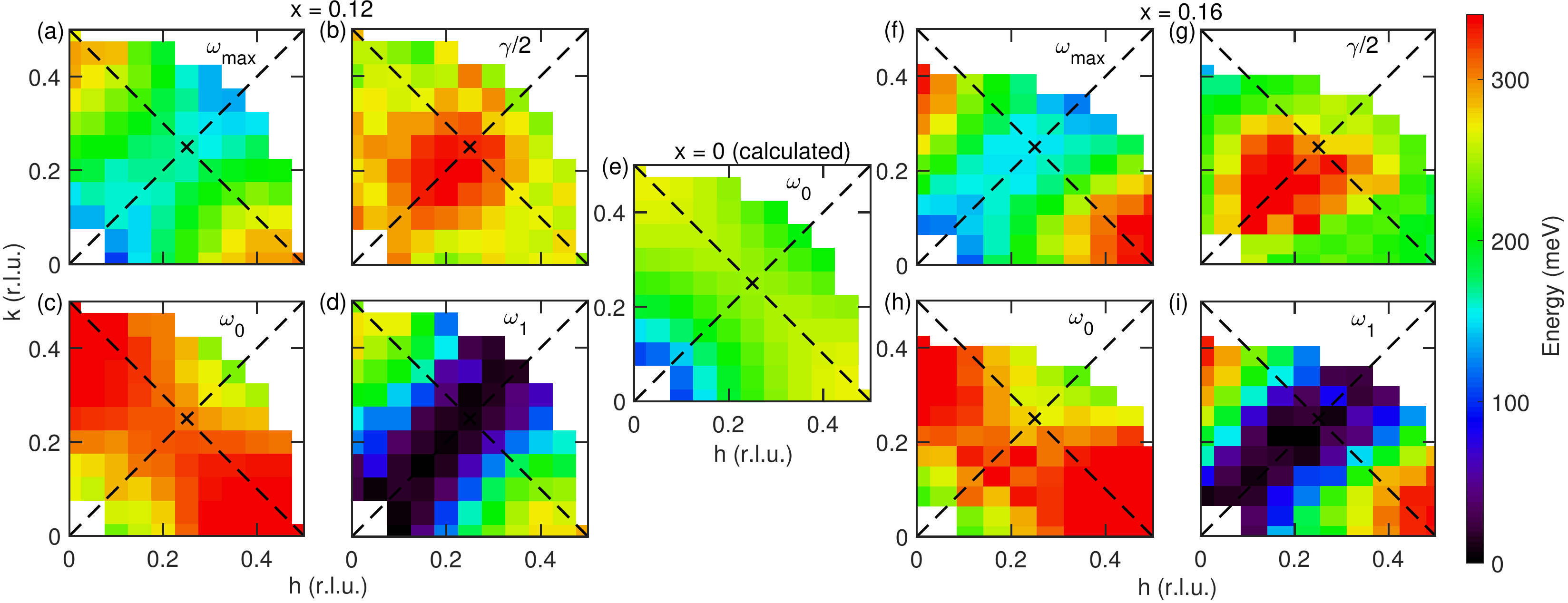}
\end{center}
\caption{Map plots showing the  parameters of a damped harmonic oscillator fit as a function of wave-vector \textbf{Q} in $x$ = 0.12 and $x$ = 0.16 LSCO. Measurements were performed at ID32 at the ESRF with $\Delta$E $\simeq$ 50\;meV. Showing the maximum of the magnon peak $\omega_{\textrm{max}}$ in (a) and (f), the damping factor $\gamma/2$ in (b) and (g) and the magnon poles $\omega_0$ in (c) and (h) and $\omega_1$ (d) and (i). The equivalent magnon pole $\omega_0$ calculated for $x$ = 0 from linear spin-wave theory with parameters from INS is shown in (e). Black dashed lines indicate the zone boundary and high symmetry directions.}
\label{fig:map_damping}
\end{figure*}

We also fit the lower resolution ($\Delta E\simeq$ 50\;meV) spectra from the grid in $(h,k)$.  The results of fitting this data to the DHO model are summarised in Fig.~\ref{fig:map_damping}.  The damping $\gamma/2$ is again seen to be largest in the region near ($0.2,0.2$) for both doped compositions.  The over-damped region where $\omega_0^2<\gamma^2/4$ and $\omega_1$ is imaginary, is indicated in Fig. \ref{fig:map_damping} (d,i) as $\omega_1=0$. 

It should be noted that normalising the data to an integration over the range of the $dd$ excitations does not account for the energy-dependence of the self-absorption. The lineshape of the excitation is therefore altered by the strong absorption of the scattered photons at low energy. We calculated that the damping parameters $\gamma$ is reduced by approximately 24\% as a result of accounting for the energy-dependent self-absorption \cite{supplementary}. This reduction decreases slightly with \textbf{Q}, and therefore the key result, that $\gamma/2$ is peaked in \textbf{Q} away from $(\frac{1}{4},\frac{1}{4})$ is unaffected.
 
%The DHO model works well for spectra measured on the doped compound, Fig. \ref{fig:fit_parameters} (a-b) shows $\omega_0$ (blue circles) and $\gamma/2$ (red squares) as a function of \textbf{Q}. The excitation energy agrees with previous work, showing the characteristic dispersion frequently reported in cuprates \cite{braicovich2009,headings2010,braicovich2010,dean2013b,meyers2017} and with the calculated excitation energy from linear spin-wave theory (SWT) (dashed blue line). Moreover, $\gamma/2$ $\ll$ $\omega_0$, indicating the excitations are underdamped. Approaching ($\frac{1}{2}$, 0), a slight increase in  $\gamma/2$ is observed, showing the enhanced damping at this position which is reported in previous work \cite{vignolle2007}.}

%As above, the damping coefficients can be extracted from the mapped data as shown in Fig. \ref{fig:map_damping}, where the fit coefficients $\omega_0$ and $\gamma/2$ are shown alongside extracted values of $\omega_1$ and $\omega_{\textrm{max}}$. Fig. \ref{fig:map_damping} (e) shows $\omega_0$ calculated for the $x$ = 0 compound from SWT as described in appendix \ref{appendix}. The calculated $\omega_0$ for the parent compound looks similar to measured $\omega_0$ in the doped compounds in Fig. \ref{fig:map_damping} (c) and (h), indicating that the undamped pole of $\omega_0$ has a similar character to spin-waves in the undoped parent compound. The distribution of $\omega_{\textrm{max}}$ in Fig. \ref{fig:map_damping} (a) and (f) is significantly more anisotropic, suggesting anisotropy is related to the emergence of damped behaviour.}

\subsection{Estimate of the absolute wavevector-dependent susceptibility}
\label{est_chi}
Fitting our RIXS data to the DHO response function in Eqn. \ref{equ:damping}, allows the wavevector-dependent susceptibility $\chi^{\prime}(\mathbf{Q})$ to be estimated, where
\begin{equation}
\label{equ:KK}
\chi^{\prime}(\mathbf{Q}) =  \chi^{\prime}(\mathbf{Q}, \omega=0) = 
\frac{1}{\pi} \int_{-\infty}^{\infty} \frac{\chi^{\prime\prime}(\mathbf{Q},\omega)}{\omega} \; d\omega.
\end{equation}
In this section, we estimate $\chi^{\prime}(\mathbf{Q})$ in the superconductors we have investigated by using the parent antiferromagnet La$_2$CuO$_4$ as a reference. This estimation assumes that the DHO response contains only magnetic contributions which is somewhat justified by recent polarisation analysis \cite{peng2018,fumagalli2019}. The data of Peng \textit{et al.} suggest that approximately 82\% of spectral weight is magnetic in the region of the magnetic excitations, 150--600\;meV. In our analysis, the 18\% charge contribution is partially accounted for in the background and multimagnon fits but any remaining charge contribution may lead to an overestimation of $\chi^{\prime}(\mathbf{Q})$.

The analysis discussed so far has relied on normalisation to $g$, an integration over the region of the $dd$ excitations, to take account of angle-dependent effects on the RIXS intensity. This does not affect the determination of excitation energies or damping coefficients.  However, this procedure does not account for the difference in absorption between photons scattered from the magnetic excitations, which are close to the resonance, and the $dd$-excitations which are significantly away from the Cu absorption peak (with a width of about 0.4\;eV) and are therefore less likely to be absorbed. In order to correct for these effects, we use the measured spin wave RIXS intensity of the parent compound as a reference. INS measurements show that the magnetic excitations of La$_2$CuO$_4$ are fairly well-described by linear spin wave theory (SWT) with some corrections\cite{headings2010} near ($\frac{1}{2},0$). Thus the underlying $S(\mathbf{Q},\omega)$ is known in this case.

Ament \textit{et al.} \cite{ament2011} point out that under certain theoretical approximations, the absolute RIXS cross-section can be split into a prefactor $f(\boldsymbol{\epsilon},\boldsymbol{\epsilon}^{\prime}, \mathbf{k}, \mathbf{k}^{\prime})$ multiplied by a dynamic structure factor $S(\mathbf{Q},\omega)$, where the polarisations of the initial and final photons are $\boldsymbol{\epsilon}$ and $\boldsymbol{\epsilon}^{\prime}$. We note that the exact circumstances when the RIXS response is proportional to $S(\mathbf{Q},\omega)$ is still an active subject of investigation \cite{ament2011,Jia2014_JNWK}, however, we will use this approximation in our analysis.  Here we propose a simple estimate to remove the effects of $f(\boldsymbol{\epsilon},\boldsymbol{\epsilon}^{\prime}, \mathbf{k}, \mathbf{k}^{\prime})$ from $S(\mathbf{Q},\omega)$ for doped LSCO. We assume $f(\boldsymbol{\epsilon},\boldsymbol{\epsilon}^{\prime}, \mathbf{k}, \mathbf{k}^{\prime})$ is the same for doped and undoped compounds. For each $(\mathbf{k}, \mathbf{k}^{\prime})$ we first normalise (divide) the raw RIXS spectra by $g$ to yield $I_{\textrm{RIXS}}$
(see Sec.~\ref{sec:methods_analysis}) and find $\chi^{\prime}_{\textrm{RIXS}}$ by fitting to the DHO model. We then multiply $\chi^{\prime}_{\textrm{RIXS}}$ for LSCO by the spin-wave response of LCO determined from INS \cite{headings2010} divided by the measured RIXS response of LCO, to estimate the dynamic susceptibility of the doped superconductor in absolute units:
\begin{equation}
\left< \chi^{\prime\,\textrm{LSCO}}(\mathbf{Q}) \right> 
= \chi^{\prime\,\textrm{LSCO}}_{\textrm{\ RIXS}}(\mathbf{Q}) \times    \frac{\phi_{\textrm{SWT}}^{\textrm{LCO}}(\mathbf{Q})}{\phi_{\textrm{RIXS}}^{\textrm{LCO}}(\mathbf{Q})}. 
\label{Eqn:normalisation}
\end{equation}
$\phi^{\textrm{LCO}}_{\textrm{SWT}}$ is the energy integrated spin-wave pole weight, determined from a fit of linear SWT to INS data and $\phi^{\textrm{LCO}}_{\textrm{RIXS}}$ is the integrated pole weight of fitted RIXS data, details of this are given in appendix \ref{appendix}. In practice, we fit the LCO spectra and then use Eqns.~\ref{eqn:LCO_SWT} and \ref{eqn:LCO_RIXS} to evaluate $\phi^{\textrm{LCO}}_{\textrm{SWT}}$ and $\phi^{\textrm{LCO}}_{\textrm{RIXS}}$. Eqn. \ref{Eqn:normalisation} assumes that the factors $f$ and $g$ are the same in doped and undoped compositions and therefore cancel in the normalisation procedure. We have verified that this is approximately the case in our samples.

\begin{figure}
\begin{center}
\includegraphics[width=\linewidth]{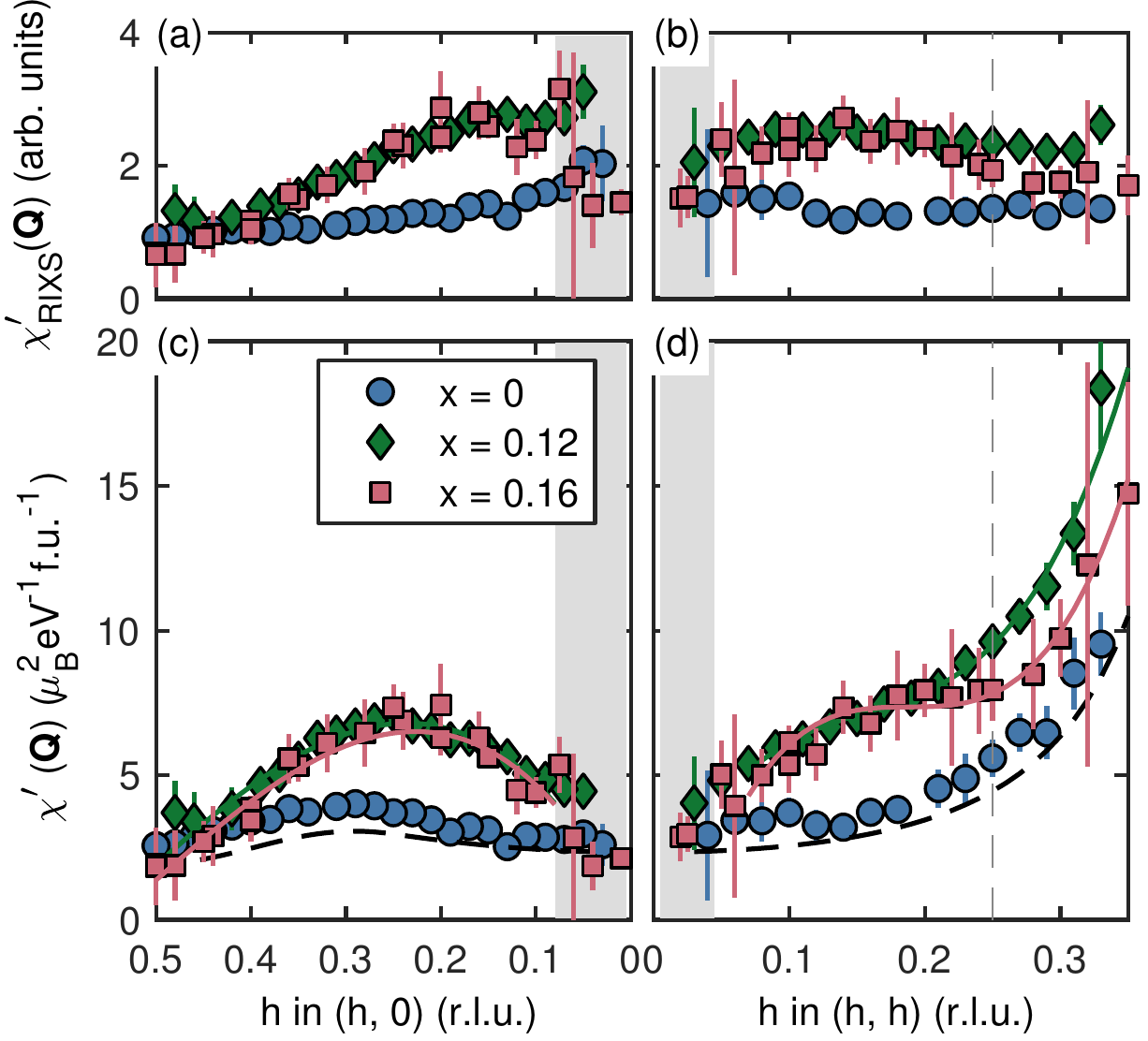}
\end{center}
\caption{Wavevector-dependent susceptibilities $\chi^{\prime}(\mathbf{Q})$ in La$_{2-x}$Sr$_x$CuO$_4$ determined from RIXS spectra. Fits of a damped harmonic oscillator function to $I_{\textrm{RIXS}}$ yield estimates of $\chi^{\prime}_{\textrm{RIXS}}(\mathbf{Q})$ shown in (a) and (b) which include self-absorption and other orientation-dependent effects. (c) and (d) show estimates of the absolute $\chi^{\prime}(\mathbf{Q})$. These estimates are obtained by normalising the data from doped compositions by the antiferromagnetic parent compound as described in the text. Cubic polynomial fits to the  data are shown as solid lines and the dashed line shows the SWT model. The dashed grey line indicates the Brillouin zone boundary. Data on all compounds were collected at I21 and additional data on the $x$ = 0.16 compound were measured at ID32.} 
\label{fig:chi_fit}
\end{figure}

Figs. \ref{fig:fit_parameters} and \ref{fig:chi_fit} (a) and (b) show the parameters $\gamma(\mathbf{Q})$, $\omega_0(\mathbf{Q})$ and $\chi^{\prime}(\mathbf{Q})$ extracted from fits of Eq.~\ref{equ:damping} as a function of \textbf{Q} along ($h, 0$) and ($h, h$) for the three compounds. For LCO, $\chi^{\prime\prime}(\textbf{Q},\omega)$ is a sum of the single and multimagnon contributions.  For LSCO a single response function is used. The resulting $\chi^{\prime}_{\text{RIXS}}(\mathbf{Q})$ due to the magnon pole is shown in 
Fig.~\ref{fig:chi_fit} (a) and (b) with a cubic polynomial fit indicated with a solid blue line.  The susceptibilities $\chi^{\prime}_{\text{RIXS}}(\mathbf{Q})$ in Fig.~\ref{fig:chi_fit}(a) and (b) contain the effects of the $f$ factor and self absorption mentioned above. In Fig.~\ref{fig:chi_fit} (c) and (d) we correct for these effects and estimate the absolute $\chi^{\prime}(\mathbf{Q})$ using Eqns.~\ref{Eqn:normalisation}, \ref{eqn:LCO_SWT} and \ref{eqn:LCO_RIXS} together with the cubic polynomial fit of $\chi^{\prime}_{\text{RIXS}}(\mathbf{Q})$ to La$_2$CuO$_4$ in Fig.~\ref{fig:chi_fit}(a,b).  
\begin{table}
\begin{center}
\begin{ruledtabular}
\begin{tabular}{c | r  r }

\textbf{Q} & $(1/4, 0)$  & $(1/4, 1/4)$ \\
\hline
$x$  &  \multicolumn{2}{c}{ $\chi^{\prime}(\mathbf{Q})   (\mu^2_{B}\textrm{eV}^{-1}\textrm{f.u.}^{-1}$ )} \\ \hline 
0 &  $3.7 \pm 0.3$  & $5.6\pm0.6$ \\
0.12  & $7.1\pm0.3$    & $9.6\pm1$ \\
0.16  &  $7.3\pm0.8$  & $8.0\pm1$  \\
\end{tabular}
\end{ruledtabular}
\end{center}
\label{tab:chi_results}
\caption{Doping dependence of the $\chi^{\prime}(\mathbf{Q})$ in LSCO as measured with RIXS.}
\end{table}

\begin{figure}
\begin{center}
\includegraphics[width=\linewidth]{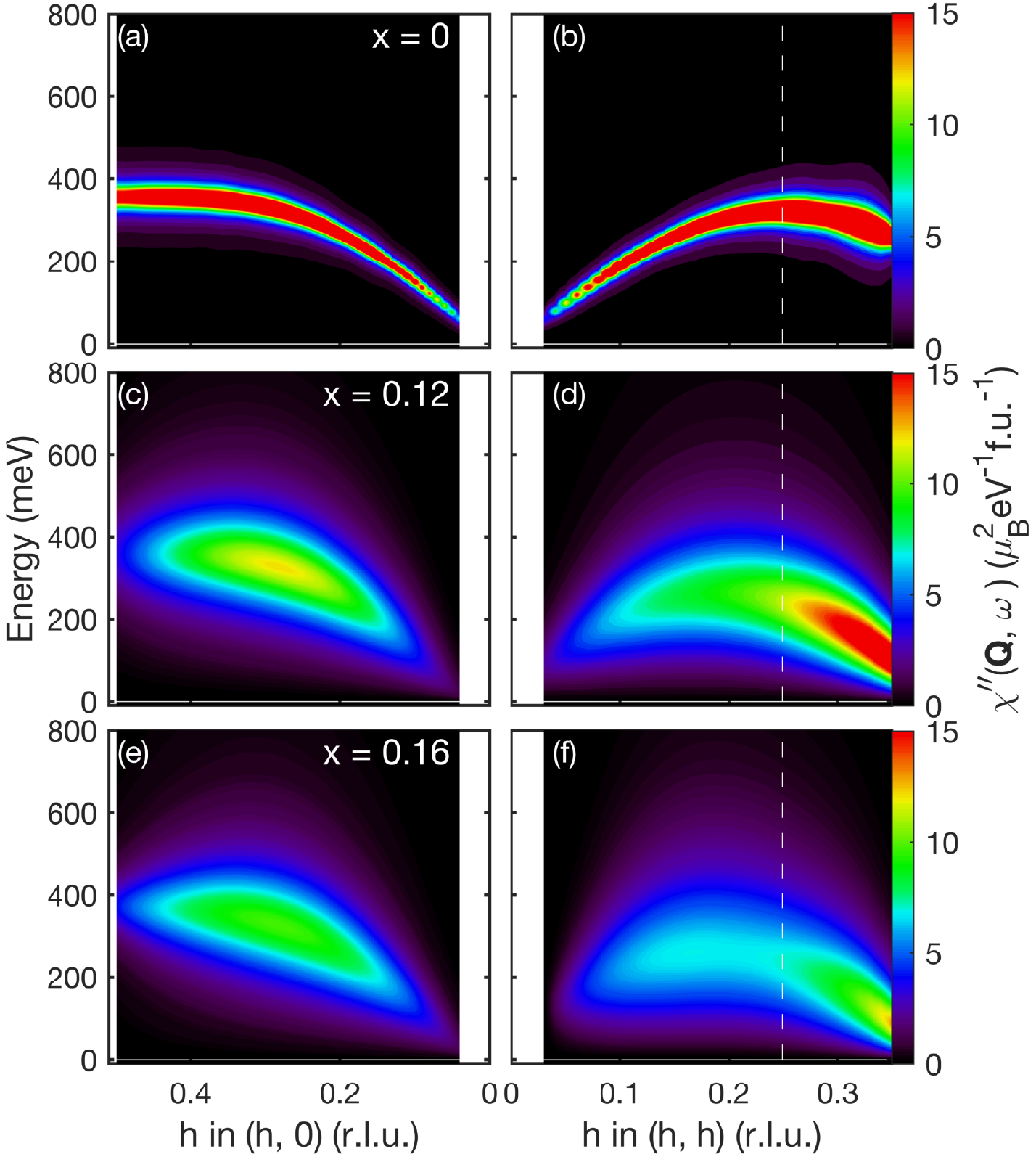}
\end{center}
\caption{(a-f) Intensity plots of $\chi^{\prime\prime}(\mathbf{Q}, \omega)$ showing the modelled excitations using the fitted parameters [$\omega_0(\mathbf{Q})$, $\gamma(\mathbf{Q})$, $\chi^{\prime}(\mathbf{Q})$] from Figs.~\ref{fig:fit_parameters} and \ref{fig:chi_fit} with Eqn.~\ref{equ:damping}. } 
\label{fig:chi_color_plot}
\end{figure}

By definition, the corrected susceptibility for the parent compound La$_2$CuO$_4$ becomes that of the SWT model described in Appendix A plus additional spectral weight due to the multimagnon excitations observed with RIXS. For all three compositions investigated, $\chi^{\prime}(\mathbf{Q})$ increases as we move  along $(h,h)$ towards $(\frac{1}{2},\frac{1}{2})$, where INS finds the strongest spin fluctuations.  The magnitude of $\chi^{\prime}(\mathbf{Q})$ is generally larger for the doped compositions $x=0.12,0.16$ than in the parent (see Table \ref{tab:chi_results}), this effect is also present when the data is normalised via the $dd$ excitations so does not seem to be an artefact arising from the spin wave normalisation. The increase arises when spectral weight in $\chi^{\prime\prime}(\mathbf{Q},\omega)$ is moved to lower energy and gives a larger contribution to $\chi^{\prime}(\mathbf{Q})$ because of the $1/\omega$ factor in Eqn.\;\ref{equ:KK}. For example, if, for a particular $\mathbf{Q}$, a spin wave keeps the same integrated intensity in $\chi^{\prime\prime}(\mathbf{Q},\omega)$ and is broadened in $\omega$, then $\chi^{\prime}(\mathbf{Q})$ can increase. Inspection of Fig.\;\ref{fig:fit_plots} shows that this indeed happens. The modelled excitations are shown in Fig. \ref{fig:chi_color_plot} where $\chi^{\prime\prime}(\mathbf{Q},\omega)$ is calculated from Eqn. \ref{equ:damping} with the fitted parameters [$\omega_0(\mathbf{Q})$, $\gamma(\mathbf{Q})$, $\chi^{\prime}(\mathbf{Q})$] shown in Figs.~\ref{fig:fit_parameters} and \ref{fig:chi_fit} (c, d).

\section{Discussion}
\subsection{Theoretical Models}
Our investigation of the magnetic excitations in cuprates is motivated by spin-fluctuation mediated theories of high temperature superconductivity\cite{scalapino2012} and to gain a fundamental understanding of metallic transition metal oxides. The Hubbard model (in its one or three band variants) is generally considered to be a good starting point. Calculations based on the Hubbard model\cite{scalapino2012} show that the wavevector-dependent pairing interaction $V_{\text{eff}}$ is approximately \cite{scalapino1995,scalapino2012}
\begin{equation}
V_{\text{eff}} \simeq \frac{3}{2} U^2 \chi^{\prime}(\mathbf{p}^{\prime}-\mathbf{p}),
\label{Eqn:pairing} 
\end{equation}
where $\mathbf{p}^{\prime}$ and $\mathbf{p}$ are the wavevectors of the two electrons making up a Cooper pair and $U$ is the Hubbard on-site interaction. RIXS measurements of the magnetic excitations over a wide energy range allow the opportunity to determine $\chi^{\prime}(\mathbf{q})$.  This can be used as an input to theory or a test of models of the excitations.  Numerical studies of the two-dimensional Hubbard model, applied to cuprates, qualitatively reproduce \cite{huang2017} the slowly-evolving high-energy magnetic excitations which are observed by INS and RIXS experiments, but calculations are restricted to relatively small lattices. Other approaches based on renormalised itinerant quasiparticles \cite{james2012,Eremin2013,dean2014,monney2016} with various types of approximation provide a basis for a phenomenological understanding of the physical properties and allow finer structure in wavevector and energy to be predicted. In general, we expect the magnetic excitations and $\chi^{\prime\prime}(\mathbf{Q},\omega)$ to be different around $(0,0)$ and $(\frac{1}{2}, \frac{1}{2})$ and the dispersion of the excitations not to be symmetric around $(\frac{1}{4},\frac{1}{4})$.   

\subsection{Wavevector dependence of the response}
\label{sec:chiq}
The high-energy magnetic excitations in the parent compound La$_2$CuO$_4$ are anisotropic in two ways. Firstly the single magnon energy varies between points on the antiferromagnetic Brillouin zone boundary with $(\frac{1}{2},0)$ having a higher energy than $(\frac{1}{4},\frac{1}{4})$. Secondly, the single magnon excitation is strongly and anomalously damped at the $(\frac{1}{2}, 0)$ position.  This variation in the magnon energy can be understood in terms of an expansion of the single band Hubbard model\cite{coldea2001,MacDonald1990} which gives rise to second nearest neighbour and cyclic exchange interactions. While the anisotropy of the damping in La$_2$CuO$_4$ may be understood in terms of the unbinding of magnons into spinons\cite{headings2010,dalla2015}. This is a generic property\cite{headings2010,dalla2015} of $S=1/2$ square lattice antiferromagnets.
 
Our data show how the anisotropies of the parent compound persist into the doped compositions and are qualitatively consistent with previous studies\cite{dean2013,monney2016,meyers2017}. However, the higher energy resolution of the present study ($\Delta E \approx 35$\;meV as compared to $\Delta E \gtrsim 100$\;meV in previous work\cite{dean2013,monney2016,meyers2017}) allows us to separate the magnetic excitations from lower energy features. In Fig.~\ref{fig:fit_parameters} we see that the frequency of the undamped mode $\omega_0(\mathbf{Q})$ extracted from the DHO model shows similar dispersions along $(h,0)$ and $(h, h)$ in the doped $x=0.12$ and $x=0.16$ compositions as in the parent $x=0$. At $\mathbf{Q}=(\frac{1}{2},0)$,  $\hbar \omega_0$ increases with doping from $356 \pm 45$\;meV $(x=0)$ to {$396 \pm 54$\;meV ($x=0.16$)}, while at $\mathbf{Q}=(\frac{1}{4},\frac{1}{4})$ it increases from $298 \pm 27$ \;meV to $313 \pm 30$\;meV. 

A new result from this work is the extent of the variation of $\gamma(\textbf{Q})$ and $\omega_0(\mathbf{Q})$ across the Brillouin zone in doped LSCO.  Significantly, the damping is seen to increase in the underdoped compound, $x$ = 0.12 and again in the optimally-doped material, $x$ = 0.16.  From the damping maps shown in Fig.~\ref{fig:map_damping} (b) and (g) it can be seen that the enhanced damping is most prominent close to the ($h, h$) direction. It is notable that the maxima in $\gamma(\textbf{Q})$ and $\omega_0(\mathbf{Q})$ along $(h,h)$ are actually near ($0.2,0.2$) rather than at ($\frac{1}{4}, \frac{1}{4}$).   Our $(h,k)$ maps of the fitted parameters in Fig.~\ref{fig:map_damping} show that $\gamma(\mathbf{Q})$ actually shows a local maximum around this point. These features appear to be qualitatively present in theoretical calculations based on itinerant quasiparticle such as those in Refs.~\onlinecite{dean2014,monney2016} and presumably arise from (nesting) features in the underlying quasiparticle band structure. The general damping anisotropy between $(h, 0)$ and $(h, h)$ for the doped compositions has also been described by theories based on determinantal quantum Monte Carlo (DQMC) \cite{huang2017}. 

The normalisation procedure described in Sec.~\ref{est_chi} allows us to obtain the estimates of $\chi^{\prime}(\mathbf{Q})$ in Fig.\;\ref{fig:chi_fit}. Values of $\chi^{\prime}$ at representative wavevectors are shown in Table \ref{tab:chi_results}. A striking feature of the analysis is that  it shows that there is a large anisotropy in $\chi^{\prime}(\mathbf{Q})$ at the antiferromagnetic Brillouin zone boundary. In particular, $\chi^{\prime}(\mathbf{Q})$ is about 4 times larger at $(\frac{1}{4},\frac{1}{4})$ than at $(\frac{1}{2},0)$.  This arises because of the smaller $\omega_0(\mathbf{Q})$ at $(\frac{1}{4},\frac{1}{4})$ (see Figs.\;\ref{fig:fit_plots} and \ref{fig:fit_parameters}) which shifts spectral weight to lower energy. 

A maximum on $\chi^{\prime}(\mathbf{Q})$ along ($h, 0$) is seen for all compositions. This may derive from the combination of two effects present in the parent antiferromagnetic state.  Firstly, linear spin-wave theory of a square lattice $S=1/2$ antiferromagnet (see Appendix A) predicts that $\chi^{\prime}(\mathbf{Q})$ increases from $(0,0)$ to $(\frac{1}{2},0)$. Secondly, square lattice $S=1/2$ antiferromagnets such as La$_2$CuO$_4$\cite{headings2010} and CFDT\cite{dalla2015} show anomalous broadening and weakening of their magnetic excitations near $(\frac{1}{2},0)$ and thus a dip in $\chi^{\prime}(\mathbf{Q})$ at this position. This is not predicted in the pure SWT model and has been understood in terms of the unbinding of magnons into spinon pairs\cite{dalla2015}. Our results suggest that these effects persist for doped compositions.

\begin{figure}
\begin{center}
\includegraphics[width=\linewidth]{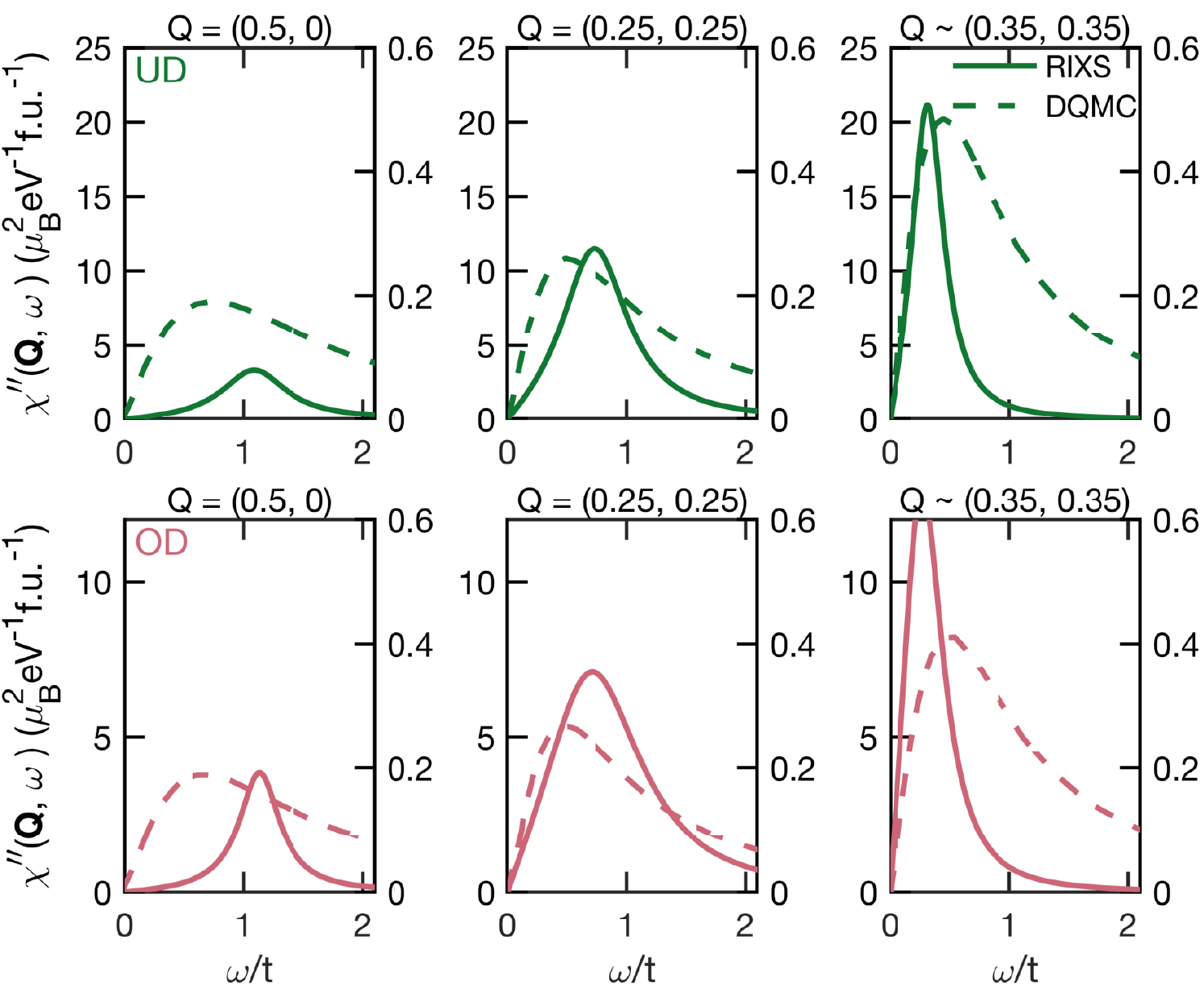}
\end{center}
\caption{Comparison of $\chi^{\prime\prime}(\mathbf{Q})$ modelled from the RIXS parameters and calculations in DQMC by Huang \textit{et al.} \cite{huang2017}. Showing our modelled spectra from the $x$ = 0.12 compound (solid green line) compared to calculations at $x$ = 0.1 (dashed green line) and spectra from the $x$ = 0.16 compound (solid pink line) compared to calculations at $x$ = 0.15 (dashed pink line). The plots are scaled differently, the RIXS scale is shown on the left axis and the DQMC scale is shown on the right.}
\label{fig:chi_slice}
\end{figure}

Also of interest is the fact that $\chi^{\prime}(\mathbf{Q})$ increases monotonically along $(h,h)$ from $\Gamma$ to $M$.  The increase is consistent with the fact that the magnetic response is strongest in the antiferromagnetic Brillouin zone centred on $M$.  This is expected because of the residual antiferromagnetic exchange interactions and is qualitatively consistent with INS measurements\cite{hayden1996,vignolle2007,Wakimoto2007_WYTF, lipscombe2007, lipscombe2009}. Thus, to our knowledge, Fig. \ref{fig:chi_fit} (c), (d) are the first attempts to determine $\chi^{\prime}(\mathbf{Q})$ in absolute units based on integrals of the magnetic response over a wide energy range. It should be noted that theoretical calculations based on the Hubbard model\cite{scalapino1995,huang2017} show that spin fluctuation in the $M$ zone contribute most to pairing in spin-fluctuation mediated theories of HTC. Fig.\;\ref{fig:chi_color_plot} shows the total modelled excitation for all compositions. The wavevector dependence of the susceptibility and damping is clearly shown.

In Fig.\;\ref{fig:chi_slice} we compare slices with calculations from the DQMC calculations of Huang et al.\cite{huang2017}. The DQMC calculations reproduce qualitatively some of the features of our data such as the increase in the strength of $\chi^{\prime\prime}(\mathbf{Q},\omega)$ moving towards $(\frac{1}{2},\frac{1}{2})$. However, the RIXS spectra are generally much sharper and show a stronger wavevector dependence.

\subsection{Comparison to INS}
%RIXS and INS provide complementary views of the collective spin excitations in the cuprates. In the antiferromagnetic parent compounds, such as La$_2$CuO$_4$ \cite{coldea2001,headings2010}, Sr$_2$CuO$_2$Cl$_2$ \cite{Plumb2014}, INS has revealed collective spin excitations up to $\sim$450\;meV with the single magnon excitation energy dispersing around the antiferromagnetic zone boundary as discussed in Sec.~\ref{sec:chiq}. The present RIXS measurements of the magnon dispersion are summarised in Fig.\;\ref{fig:fit_parameters}, where it can be seen that the magnon energy $\hbar\omega$ measured by RIXS may be slightly higher than that measured by INS. Although there are large error bars in the RIXS measurement at large wavevectors due to the difficulty in establishing the zero energy position. Thus, we find at \textbf{Q} = ($\frac{1}{2}$, 0) $\hbar \omega_0$ = $355 \pm 25$\;meV with RIXS, compared to $\hbar \omega_0$ = $322 \pm 6$~meV measured with INS\cite{headings2010}. A similar difference\cite{Plumb2014} is also suggested by data collected on Sr$_2$CuO$_2$Cl$_2$.  A difference in pole energy for RIXS and INS may be due to  different sensitivities to the fundamental transverse $\chi^{\prime\prime}_{\perp}(\mathbf{Q},\hbar \omega)$  and longitudinal $\chi^{\prime\prime}_{\parallel}(\mathbf{Q},\hbar \omega)$ components of magnetic response of an antiferromagnet\cite{lorenzana2005,dalla2015}. It is known that $\chi^{\prime\prime}_{\parallel}(\mathbf{Q},\hbar \omega)$ is peaked at higher energy in the sister $S=\frac{1}{2}$ system Cu(DCOO)$_2\cdot$4D$_2$O. 

RIXS and INS provide complementary views of the collective spin excitations in the cuprates \cite{Wakimoto2007_WYTF}. However, INS measurements of the high-energy magnetic excitations are difficult because the background increases when high incident energies are used. Nevertheless, some data does exist for La$_{2-x}$Sr$_{x}$CuO$_4$. An early study\cite{hayden1996} on La$_{1.86}$Sr$_{0.14}$CuO$_4$ revealed magnetic excitations up to 260\;meV. In particular, excitations were observed at $ \textbf{Q} = (\frac{3}{2}, 0)$ which is equivalent to the $\textbf{Q} = (\frac{1}{2},0)$ position investigated here with RIXS.  Our RIXS normalisation procedure (Sec.~\ref{est_chi}) allows us to estimate $\chi^{\prime}(\mathbf{Q})=1.8 \pm 0.6 \mu_B^2$\;eV$^{-1}$\;f.u.$^{-1}$ in LSCO $x$ = 0.16 at $(\frac{1}{2},0)$ based on an integration of the spectrum up to about 800~meV.  Integrating the INS data in Ref.~\onlinecite{hayden1996} up to 260\;meV we obtain $\chi^{\prime}(\mathbf{Q}) \approx$   0.5\;$\mu_B^2$\;eV$^{-1}$\;f.u.$^{-1}$. Thus, if it were possible to perform neutron scattering experiments over a wider energy range the integration of INS data may produce a comparable value for $\chi^{\prime}(\mathbf{Q})$ at $\textbf{Q} = (\frac{1}{2},0)$. The approximate agreement is satisfying, however further work is required to develop the comparison of the two probes of collective magnetic excitations. 

The INS study in Ref.\;\onlinecite{hayden1996} [Fig.\;4(d)] also estimated $\chi^{\prime}(\mathbf{Q})$ along the line $(h,h)$ for La$_{1.86}$Sr$_{0.14}$CuO$_4$. Unfortunately, the energy integration was only carried out over the range $0 \le \hbar \omega \le 150$\;meV. However, the increase in $\chi^{\prime}(\mathbf{Q})$ in the doped compound in the range $0.25 \le h \le 0.34$ [Fig.\;\ref{fig:chi_fit}(d), present paper] is also seen with INS. The absolute values of $\chi^{\prime}(\mathbf{Q})$ measured with neutrons are of the same order of magnitude but less than those reported in the present RIXS study presumably because the INS study integrates only up to 150\;meV.

\section{Summary and Conclusions}
We have made high-resolution RIXS measurements of the collective magnetic excitations for three compositions of the superconducting cuprate system La$_{2-x}$Sr$_{x}$CuO$_4$. Specifically, we have mapped out the excitations throughout the 2-D $(h,k)$ Brillouin zone to the extent that is possible at the Cu-$L$ edge. In addition, we have attempted to determine the wavevector-dependent susceptibility of the doped compositions  La$_{2-x}$Sr$_{x}$CuO$_4 (x=0.12,0.16)$ by normalising data to the parent compound.  This procedure allows comparison with INS measurements. We find that the evolution of the intensity of high-energy ($\hbar \omega \gtrsim 200$\;meV) excitations measured by RIXS and INS is consistent.   

The high-energy spin fluctuations in La$_{2-x}$Sr$_{x}$CuO$_4$ are fairly well-described by a damped harmonic oscillator model. The DHO damping parameter increases with doping and is largest along the ($h, h$) line although it is not peaked at the high symmetry point $(\frac{1}{4},\frac{1}{4})$. While the pole frequency is peaked at $(\frac{1}{2},0)$ for doped and undoped compositions, for the doped compositions, the wavevector-dependent susceptibility $\chi^{\prime}(\mathbf{Q})$ is much larger at $(\frac{1}{4},\frac{1}{4})$ than at $(\frac{1}{2},0)$. Both of these positions are on the antiferromagnetic zone boundary of the parent compound.  The wavevector-dependent susceptibility increases rapidly along the $(h,h)$ line towards the antiferromagnetic wavevector of the parent compound $(\frac{1}{2},\frac{1}{2})$. Thus the strongest magnetic excitations and those predicted to favour superconductive pairing occur towards the $(\frac{1}{2},\frac{1}{2})$ position. Our quantitative determination of the wavevector-dependent susceptibility will be useful in testing magnetic mediated theories of high-temperature superconductivity\cite{chubukov2003,scalapino2012}.

\appendix
\section{Linear spin-wave theory calculations}
\label{appendix}
The magnetic excitations can be modelled in LCO with classical linear spin-wave theory. We consider the case of a $S=1/2$ square lattice antiferromagnet with nearest- and next-nearest exchange interactions.  The susceptibility transverse to the ordered moment $\chi^{\prime\prime}_{\perp}(\mathbf{Q},\omega)$ due to one-magnon creation is given by:
\begin{align}
\chi^{\prime\prime}_{\perp}(\mathbf{Q},\omega) &= Z_d(\mathbf{Q}) \frac{\pi}{2} g^2 \mu_B^2 S \left( \frac{A_{\mathbf{Q}}-B_{\mathbf{Q}}}{A_{\mathbf{Q}}+B_{\mathbf{Q}}} \right)^{1/2} \! \!\delta[\omega \pm \omega_0(\mathbf{Q})] \nonumber  \\
&= \frac{\pi}{2} \chi^{\prime}_{\perp}(\mathbf{Q}) \, \omega_0(\mathbf{Q}) \, \delta[\omega \pm \omega_0(\mathbf{Q})], \label{chi_im_perp}
\end{align}
where   
\begin{equation} 
\hbar \omega_0(\textbf{Q})  =  2Z_c \sqrt{A_{\mathbf{Q}}^2 - B_{\mathbf{Q}}^2},
\end{equation}
and
\begin{equation}
\chi^{\prime}_{\perp}(\mathbf{Q}) = Z_d(\mathbf{Q}) \frac{g^2 \mu_B^2 S}{A_{\mathbf{Q}} + B_{\mathbf{Q}}}. \label{chi_real_perp}
\end{equation}
The amplitude factors $A_{\mathbf{Q}}$ and $B_{\mathbf{Q}}$ are given\cite{coldea2001} by $A_{\mathbf{Q}} = J-J_c/2-(J^{\prime}-J_c/4)(1-v_hv_k) - J^{\prime\prime}(1 - (v_{2h} + v_{2k})/2$),
$B_{\mathbf{Q}} = (J-J_c/2)(v_h+v_k)/2$,
where $v_x = \cos(2\pi x)$ and $x \mapsto h$ or $k$. $Z_d$ and $Z_c$ are renormalisation constants which take account of quantum fluctuations in the AF ground state. 

Headings \textit{et al.} \cite{headings2010} have made INS measurements of the spin waves in La$_2$CuO$_4$ and fitted the model described by Eqns.~\ref{chi_im_perp}-\ref{chi_real_perp}. They find
$J$ = 143, $J^{\prime}$ = $J^{\prime\prime}$ = 2.9 and $J_c$ = 58~meV, assuming $Z_c=1.18$. The wavevector dependence of $Z_d(\mathbf{Q})$ is also determined from the INS data,
\begin{equation}
Z_d(\mathbf{Q}) = \begin{cases} \mbox{$Z_{d0}\sin(h\pi)$,} & \mbox{if } h < \frac{1}{4} \\ \mbox{$Z_{d0}\sin({\displaystyle \frac{\pi}{4}})$,} & \mbox{if } h \geq \frac{1}{4} \end{cases}, 
\end{equation}
where $Z_{d0}$ = 0.4. In order to compare the INS and RIXS measurements, we assume that RIXS is equally sensitive to the three components of the susceptibility and compute the average susceptibility $\chi=\frac{1}{3}(\chi_{xx}+\chi_{yy}+\chi_{zz})=\frac{2}{3}\chi_{\perp}$.  The energy integrated intensity of the spin wave pole $\phi_{\textrm{SWT}}^{\textrm{LCO}}(\mathbf{Q})$ is then:
\begin{equation}
\phi_{\textrm{SWT}}^{\textrm{LCO}}(\mathbf{Q}) = \int_0^{\infty} \chi^{\prime\prime} (\mathbf{Q},\omega) \, d\omega
= \frac{\pi}{3} \chi^{\prime}_{\perp}(\mathbf{Q})  \, \omega_0(\mathbf{Q}).
\label{eqn:LCO_SWT}
\end{equation}
We derive a comparable measure of the energy integrated spin-wave pole measured with RIXS by rewriting Eqn.~\ref{equ:damping} for LCO (in the limit $\omega_0 \ge \gamma/2$) as,
\begin{align}
& \chi^{\prime\prime}(\mathbf{Q},\omega) = \frac{\chi^{\prime}(\mathbf{Q})}{2 \omega_1(\mathbf{Q})} \left[\frac{\gamma ^2(\mathbf{Q})}{4}+\omega_1^2(\mathbf{Q})\right] 
\times \\  &\left\{\frac{\gamma(\mathbf{Q})/2 }{ \gamma ^2(\mathbf{Q})/4+[\omega -\omega_1(\mathbf{Q})]^2}-\frac{\gamma(\mathbf{Q})/2 }{ \gamma^2(\mathbf{Q})/4+[\omega +\omega_1(\mathbf{Q})]^2}\right\}.
\end{align}
Integrating over the positive energy pole, we obtain the measured pole intensity from the fitted parameters $\omega(\textrm{Q}), \gamma(\textrm{Q})$ and $\chi^{\prime}(\textrm{Q})$:
\begin{equation}
\phi_{\textrm{RIXS}}^{\textrm{LCO}}(\mathbf{Q})=\frac{\pi \chi^{\prime}(\mathbf{Q}) \omega_0^2(\mathbf{Q}) }{\sqrt{4 \omega_0^2(\mathbf{Q})-\gamma^2(\mathbf{Q})}}.\\
\label{eqn:LCO_RIXS}
\end{equation}

\acknowledgments{The authors acknowledge funding and support from the Engineering and Physical Sciences Research Council (EPSRC) Centre for Doctoral Training in Condensed Matter Physics (CDT-CMP), Grant No. EP/L015544/1 as well as Grant EP/R011141/1. We acknowledge Diamond Light Source for time on Beamline I21 under proposals SP18469 and SP18512 and the European Synchrotron Radiation Facility for time on Beamline ID32 under proposal HC/2696. We would like to thank G. B. G. Stenning and D. W. Nye for help on the Laue instrument in the Materials Characterisation Laboratory at the ISIS Neutron and Muon Source.}

\end{document}

\title{Supplementary material for anisotropic damping and wavevector dependent susceptibility of the spin fluctuations in  La$_{2-x}$Sr$_{x}$CuO$_4$ studied by resonant inelastic x-ray scattering}
\author{H. C. Robarts} 
\affiliation{H. H. Wills Physics Laboratory, University of Bristol, Bristol BS8 1TL, United Kingdom}
\affiliation{Diamond Light Source, Harwell Science \& Innovation Campus, Didcot, Oxfordshire OX11 0DE, United Kingdom}
\author{M. Barth\'elemy}
\affiliation{H. H. Wills Physics Laboratory, University of Bristol, Bristol BS8 1TL, United Kingdom}
\author{K. Kummer}
\affiliation{European Synchrotron Radiation Facility, 71 Avenue des Martyrs, Grenoble, France}
\author{M. Garc\'ia-Fern\'andez}
\affiliation{Diamond Light Source, Harwell Science \& Innovation Campus, Didcot, Oxfordshire OX11 0DE, United Kingdom}
\author{J.\;Li}
\affiliation{Diamond Light Source, Harwell Science \& Innovation Campus, Didcot, Oxfordshire OX11 0DE, United Kingdom}
\affiliation{Beijing National Laboratory for Condensed Matter Physics and Institute of Physics, Chinese Academy of Sciences, Beijing 100190, China}
\author{A. Nag}
\author{A. C. Walters}
\author{K. J. Zhou}
\email{kejin.zhou@diamond.ac.uk}
\affiliation{Diamond Light Source, Harwell Science \& Innovation Campus, Didcot, Oxfordshire OX11 0DE, United Kingdom}
\author{S. M. Hayden}
\email{s.hayden@bristol.ac.uk}
\affiliation{H. H. Wills Physics Laboratory, University of Bristol, Bristol BS8 1TL, United Kingdom}

\maketitle
\section{Self-absorption correction}
In our initial data analysis we normalise the data to an integration over the range of the $dd$ excitations yielding an approximation of the true RIXS intensity $I_{\textrm{RIXS}}$. This estimation should account for the geometrical effect of self-absorption but does not consider its energy dependence. In fact, the energy dependence of the self-absorption is significant at low energy, especially when the energy of the incident photons is tuned to the peak of the absorption edge, as is the case in our experiments.

The energy dependent self-absorption slightly alters the lineshape of the low-energy excitations and hence affects our estimation of the damping parameter $\gamma$. There is no affect on our estimation of $\chi^{\prime}(\mathbf{Q})$ as the energy dependence of the self-absorption is accounted for in Eqn. 7 of the main paper.

To understand the effect on the excitation lineshapes, we estimate the self-absorption on the $x = 0.12$ data and compare the fitted $\gamma$ parameters for each normalisation method. The intensity of the scattered photons is assumed to be reduced by a self-absorption factor which is dependent on the incident photon angle $\theta$ and total scattering angle $\Omega$ and x-ray absorption coefficients which are defined as $\mu_i$ and $\mu_f$ for the incident and final scattered photons respectively,
\begin{equation}
I \propto I \frac{\sin(\Omega - \theta)}{\mu_i\sin(\Omega - \theta) + \mu_f\sin\theta}.
\label{eqn:self_absorption}
\end{equation}
The absorption coefficients $\mu_i$ and $\mu_f$ are dependent on the energy, polarisation and geometry of the incident and scattered light. We follow recent practice \cite{comin2015, comin2015b, minola2015, achkar2016,fumagalli2019} by estimating the absorption coefficients from x-ray absorption spectroscopy (XAS) which we performed prior to the RIXS measurements at I21 by measuring the total electron yield from the sample. We assume that scattered photons are flipped from $\pi$ to $\sigma$ polarisation, however, as we do not perform polarisation analysis on the scattered photons, the exact self-absorption cannot be fully accounted for \cite{fumagalli2019}.

\begin{figure*}
\begin{center}
\includegraphics[width=\linewidth]{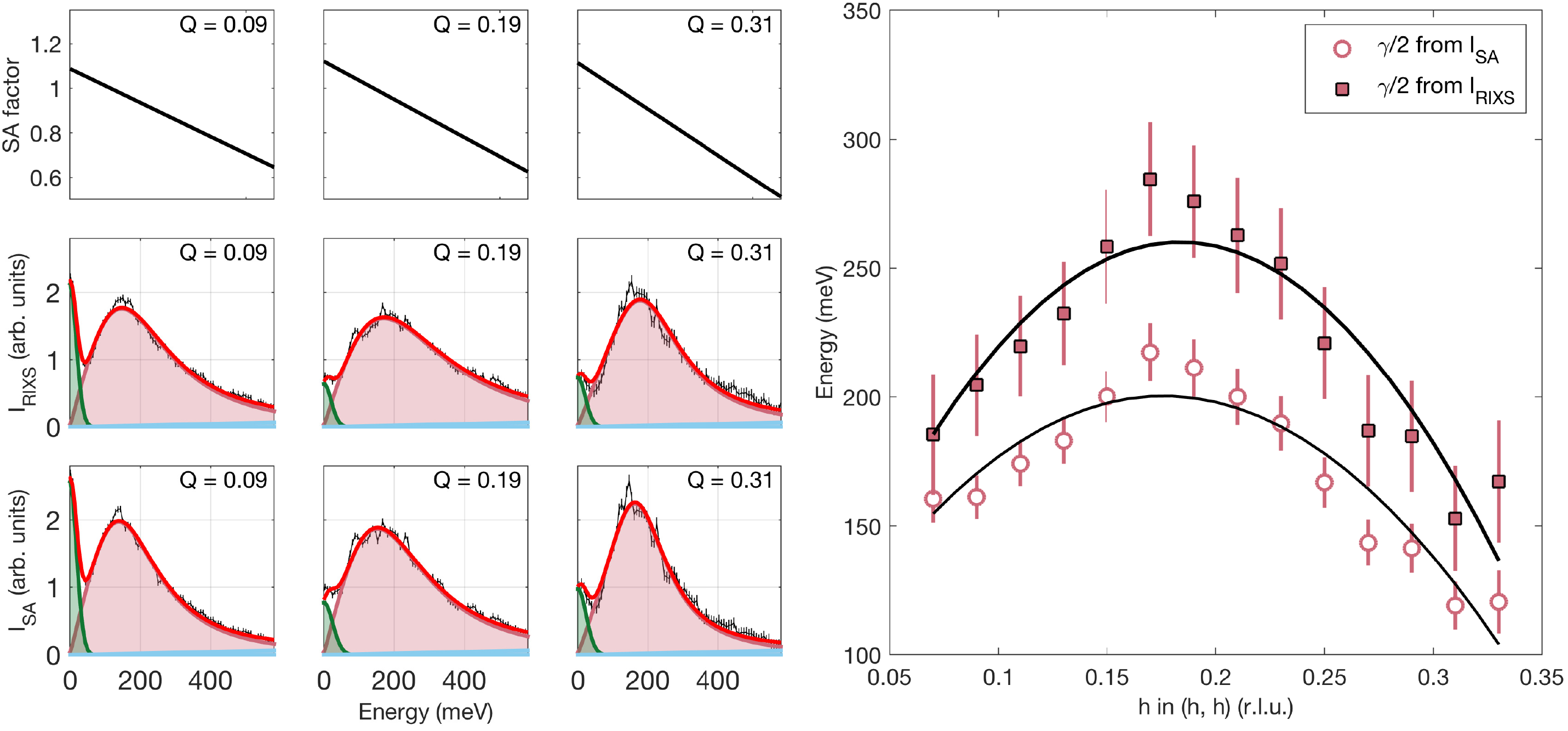}
\end{center}
\caption{Calculated self absorption factors as a function of energy for $\pi$ to $\sigma$ polarisation at \textbf{Q} = (0.09, 0.09), (0.19, 0.19) and (0.31, 0.31). Also showing a comparison of spectral lineshape for LSCO $x$ = 0.12 data normalised to the $dd$ excitations, shown as $I_{\textrm{RIXS}}$ and data corrected for the effects of self absorption shown as $I_{\textrm{SA}}$. The fitted damping parameter $\gamma/2$ is compared for the two normalisation methods.}
\label{fig:self_absorption}
\end{figure*}

Fig. \ref{fig:self_absorption} shows the calculated self-absorption factors for three wavevectors and the resulting excitation lineshape $I_{\textrm{SA}}$ compared to the approximation used in the main paper $I_{\textrm{RIXS}}$. Both plots show the data fit to the damped harmonic oscillator model and the damping parameter $\gamma$ is plotted in the right panel. $\gamma$ extracted from the self-absorption corrected data is an average of 24\% lower. However, the wavevector dependent anisotropy is still clearly evident with a peak centred at \textbf{Q} $\simeq 0.19$.

%\bibliographystyle{apsrev4-1}
%\bibliography{../../../Papers/bibpapers}
%\bibliography{bibpapers}

%